\documentclass[twocolumn]{aastex631}
\usepackage{threeparttable}

\shorttitle{A Barred Spiral DSFG}
\shortauthors{Huang et al.}
\graphicspath{{./}{figures/}}

\begin{document}
\title{J0107a: A Barred Spiral Dusty Star-forming Galaxy at $z=2.467$}

\author[0009-0006-1731-6927]{Shuo Huang}
\affiliation{Institute of Astronomy, Graduate School of Science, The University of Tokyo, 2-21-1 Osawa, Mitaka, Tokyo 181-0015, Japan}
\author[0000-0002-8049-7525]{Ryohei Kawabe}
\affiliation{National Astronomical Observatory of Japan, 2-21-1 Osawa, Mitaka, Tokyo 181-8588, Japan}
\affiliation{Department of Astronomy, School of Science, The Graduate University for Advanced Studies (SOKENDAI), Osawa, Mitaka, Tokyo 181-8588, Japan}
\author[0000-0002-4052-2394]{Kotaro Kohno}
\affiliation{Institute of Astronomy, Graduate School of Science, The University of Tokyo, 2-21-1 Osawa, Mitaka, Tokyo 181-0015, Japan}
\affiliation{Research Center for the Early Universe, Graduate School of Science, The University of Tokyo, 7-3-1 Hongo, Bunkyo-ku, Tokyo 113-0033, Japan}

\author[0000-0002-2501-9328]{Toshiki Saito}
\affiliation{National Astronomical Observatory of Japan, 2-21-1 Osawa, Mitaka, Tokyo 181-8588, Japan}

\author[0000-0002-6068-8949]{Shoichiro Mizukoshi}
\affiliation{Institute of Astronomy, Graduate School of Science, The University of Tokyo, 2-21-1 Osawa, Mitaka, Tokyo 181-0015, Japan}

\author[0000-0002-2364-0823]{Daisuke Iono}
\affiliation{National Astronomical Observatory of Japan, 2-21-1 Osawa, Mitaka, Tokyo 181-8588, Japan}
\affiliation{Department of Astronomy, School of Science, The Graduate University for Advanced Studies (SOKENDAI), Osawa, Mitaka, Tokyo 181-8588, Japan}

\author[0000-0003-2475-7983]{Tomonari Michiyama}
\affiliation{Faculty of Welfare and Information, Shunan University, 43-4-2, Gakuendai, Shunan, Yamaguchi, 745-8566, Japan}
\author[0000-0003-4807-8117]{Yoichi Tamura}
\affiliation{Department of Physics, Graduate School of Science, Nagoya University, Furocho, Chikusa, Nagoya 464-8602, Japan}

\author[0000-0003-4073-3236]{Christopher C. Hayward}
\affiliation{Center for Computational Astrophysics, Flatiron Institute,
162 Fifth Avenue, New York, NY 10010, USA}

\author[0000-0003-1937-0573]{Hideki Umehata}
\affiliation{Institute for Advanced Research, Nagoya University, Furocho, Chikusa, Nagoya 464-8602, Japan}
\affiliation{Department of Physics, Graduate School of Science, Nagoya University, Furocho, Chikusa, Nagoya 464-8602, Japan}
\affiliation{Cahill Center for Astronomy and Astrophysics, California Institute of Technology, MS 249-17, Pasadena, CA 91125, USA}

\begin{abstract}
Dusty Star-Forming Galaxies (DSFGs) are amongst the most massive and active star-forming galaxies during the cosmic noon. Theoretical studies have proposed various formation mechanisms of DSFGs, including major merger-driven starbursts and secular star-forming disks. Here, we report J0107a, a bright ($\sim8$ mJy at observed-frame 888 $\mu$m) DSFG at $z=2.467$ that appears to be a gas-rich massive disk and might be an extreme case of the secular disk scenario. J0107a has a stellar mass $M_\star\sim5\times10^{11}M_\odot$, molecular gas mass $M_\mathrm{mol}\sim(1\textendash6)\times10^{11}M_\odot$, and a star formation rate (SFR) of $\sim500M_\odot$ yr$^{-1}$. J0107a does not have a gas-rich companion. The rest-frame 1.28 $\mu$m JWST NIRCam image of J0107a shows a grand-design spiral with a prominent stellar bar extending $\sim15$ kpc. ALMA band 7 continuum map reveals that the dust emission originates from both the central starburst and the stellar bar. 3D disk modeling of the CO(4-3) emission line indicates a dynamically cold disk with rotation-to-dispersion ratio $V_\mathrm{max}/\sigma\sim8$. The results suggest a bright DSFG may have a non-merger origin, and its vigorous star formation may be triggered by bar and/or rapid gas inflow.
\end{abstract}

\section{Introduction}\label{sec:intro}
The cosmic star formation rate density increases from early times to its peak at $z\sim2-3$, called cosmic noon, and then decreases steadily to the present day \citep[for a review, see][]{2014ARA&A..52..415M}. During cosmic noon, dusty star-forming galaxies (DSFGs) populate the massive and high-SFR end of star-forming galaxies, with $M_\star\sim10^{11}M_\odot$ and SFR$\sim100-1000M_\odot$ yr$^{-1}$\citep[e.g.,][]{2015ApJ...806..110D,2017A&A...606A..17M,2020MNRAS.494.3828D}. The large infrared luminosity of DSFGs makes them easy to be detected with submillimeter/millimeter observations, and they are often referred to as submillimeter galaxies \citep[SMGs; e.g.,][]{1997ApJ...490L...5S,1998Natur.394..241H,2014PhR...541...45C}
\par
The origin of DSFGs has been debated since their discovery. Based on insights from studying local ultraluminous infrared galaxies (ULIRGs), which are exclusively late-stage major mergers \citep[e.g.,][]{1996ARA&A..34..749S}, and observations of high-z DSFGs, many authors have argued that bright DSFGs (or classical SMGs with $S_{\mathrm{850}\mu\mathrm{m}}\gtrsim5$ mJy) are mainly major mergers \citep[e.g.,][]{2008ApJ...680..246T,2010ApJ...724..233E,2022ApJ...929..159C}.
Early theoretical models demonstrated that major mergers can indeed power bright DSFGs \citep[e.g.,][]{2010MNRAS.401.1613N,2011ApJ...743..159H}, but it was unclear whether there were enough major mergers to account for the observed number counts of bright DSFGs. \citet{2011ApJ...743..159H,2012MNRAS.424..951H,2013MNRAS.428.2529H} argued that the high-z bright DSFG population is heterogeneous, with merger-driven starbursts dominating at the bright end and `quiescently star-forming' disks dominating at fainter submm fluxes.
Others have argued that the DSFG population is predominantly gas-rich massive disks (i.e. not ongoing major mergers) that sustain high SFRs due to high gas accretion rates from the cosmic web \citep[e.g.,][]{2009Natur.457..451D,2010MNRAS.404.1355D,2015Natur.525..496N,2021MNRAS.502..772L}.
\par
Morphological information can thus place critical constraints on the formation channels of DSFGs by distinguishing disks and mergers. Previous studies using deep Hubble Space Telescope (HST) imaging have revealed a prevalence of disturbed morphology \citep[e.g., ][]{2015ApJ...799..194C}, inline with interaction or merger. However, due to the high redshift and heavy dust obscuration of DSFGs, the rest-frame UV/optical emission observed with HST might not be an ideal tracer of stellar distribution.
\par
The advent of {\it James Webb} Space Telescope (JWST) makes rest-frame optical-to-NIR imaging with unprecedented sensitivity and spatial resolution available for DSFGs. Recent JWST studies of submm-selected galaxies have found a large fraction of isolated disks \citep{2022ApJ...939L...7C,2023ApJ...942L..19C}, in contrast to more clumpy and irregular shapes in HST bands. JWST rest-frame NIR imaging will provide a more robust view of the stellar distribution of DSFGs and hints for their possibly diverse formation channels.
\par
In this paper, we report J0107a, a $z=2.467$ DSFG with a clear barred spiral structure in JWST/NIRCam images and bright dust continuum and CO/[CI](1-0) line emission. We describe the observation in Section \ref{sec:data} and data analyses and results in Section \ref{sec:analysis}. We discuss the results and summarize in Section \ref{sec:summary}. Throughout this paper we assume the \cite{2003PASP..115..763C} IMF and Planck 2018 cosmology \citep[$H_0=67.4$ km s$^{-1}$ Mpc$^{-1}$, $\Omega_{m}=0.315$ and $\Omega_{\mathrm{\Lambda}}=0.685$; ][]{2020A&A...641A...1P}.
\section{DATA}\label{sec:data}
\subsection{The Target}\label{subsec:j0107a}
The target of this study, J0107a was first identified by \citet{2014ApJ...781L..39T} as a serendipitously detected line emitter at the observed frequency of 99.75 GHz in ALMA band 3 observation towards the local galaxy merger VV114 at $z=0.0205$. Based on photometric redshift, the line is identified as CO(3-2) at $z=2.467$. The redshift is confirmed by \citet{2021ApJ...917...94M} with CO(3-2), CO(4-3), and [CI](1-0) emission lines (Figure \ref{fig:fig1}a). J0107a is also detected in continuum with flux densities of $3.4\pm0.3$ mJy at 1.3 mm and $7.9\pm1.0$ mJy at $888$ $\mu$m, which satisfy the selection criteria of a classical SMG.
\subsection{ALMA Data}\label{subsec:alma}
\begin{figure*}[!ht]
\gridline{\fig{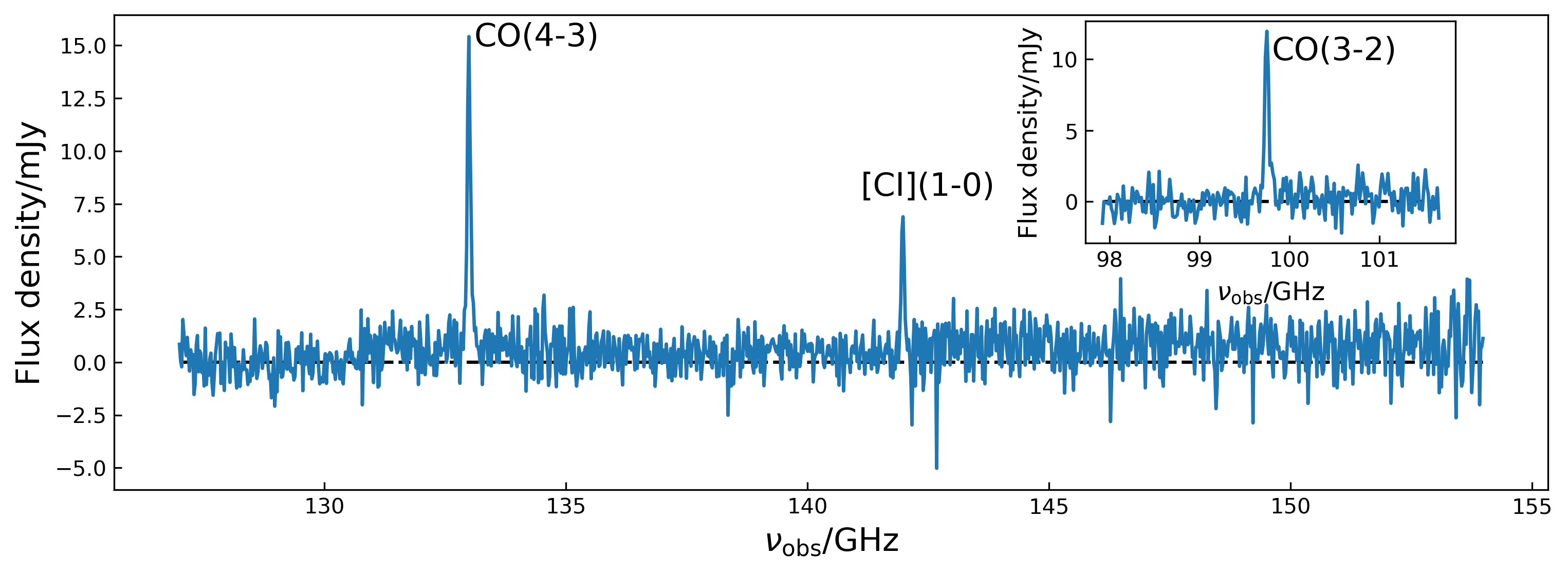}{1\textwidth}{(a)}}
    \gridline{\fig{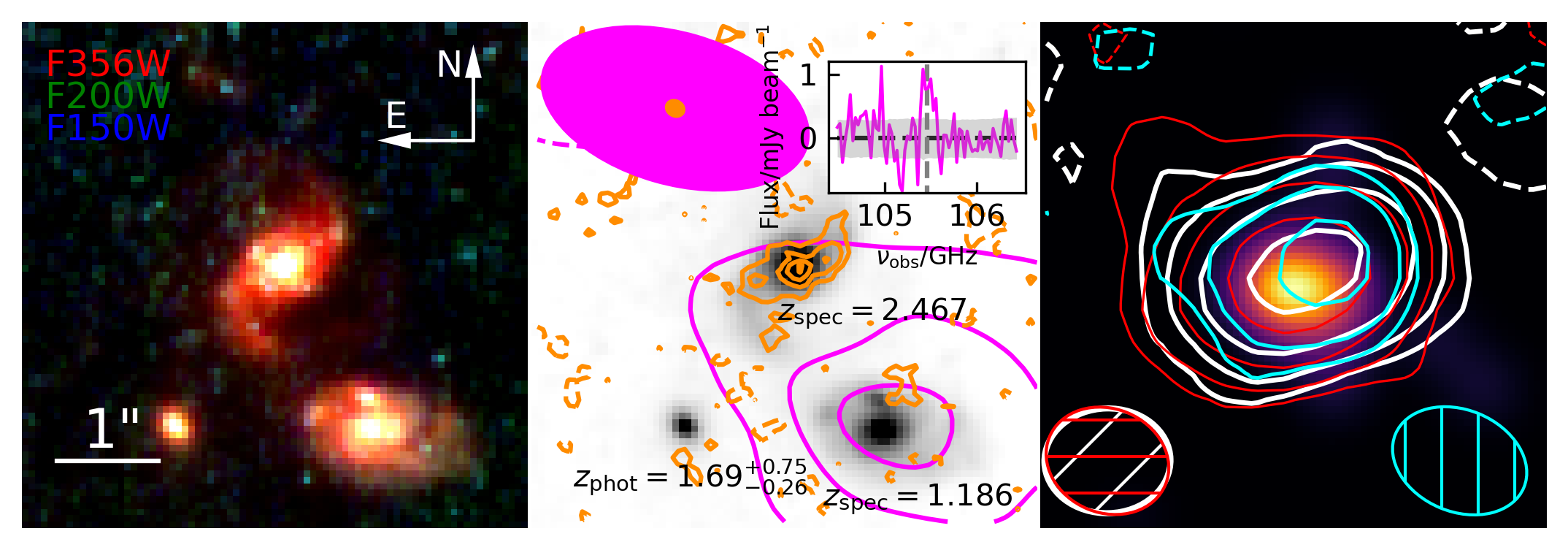}{1\textwidth}{(b)}}
    \caption{A multiwavelength view of J0107a: (a) the integrated 1D spectra of J0107a from ALMA Band 3/4 data. The channel widths are 50 km s$^{-1}$. (b) Left: JWST/NIRCam color composite image (red: F356W; green: F200W; blue: F150W) of J0107a. Middle: F356W image with ALMA band 7 continuum of J0107a (dark orange) and the zeroth-moment map of the band 3 emission line from the $z=1.186$ southwest (SW) spiral galaxy (magenta) overlaid. The insert panel shows the band 3 spectrum of the SW spiral galaxy, with the line's central frequency of 105.458 GHz marked with the gray dashed line. Right: Chandra ACIS-S 0.5-10 keV count image smoothed with a $0\farcs5$ FWHM circular Gaussian kernel. Contours show the zeroth-moment maps of J0107a's emission lines, with white, red, and blue for CO(3-2), CO(4-3), and [CI](1-0), respectively. Each panel is $5\arcsec\times5\arcsec$ in size. ALMA beam sizes are shown with the same colors as the contours. Contour levels are $-2$, 2, 4, 6, 8, 10... in the middle panel and $-2$, 2, 4, 8, 16... in the right panel.}
    \label{fig:fig1}
\end{figure*}

J0107a has been covered by the ALMA campaigns toward VV114. In this study, we include ALMA band 3 (Project code 2013.1.00469.S and 2013.1.01057.S), band 4 (2013.1.01057.S), band 6 (2015.1.00973.S and 2015.1.00902.S), and band 7 (2013.1.00740.S) data. Data reduction and imaging are conducted with CASA \citep{2022PASP..134k4501C} in a standard manner following \citet{2021ApJ...917...94M}, except for band 3 and 4, additional spectral cubes with natural weighting are made to maximize sensitivity.  The integrated 1D ALMA spectra of J0107a are shown in Figure \ref{fig:fig1}a.
\subsection{JWST Data}\label{subsec:jwst}
JWST observations of the VV114 field have been conducted as part of the GOALS-JWST survey \citep{2022ApJ...940L...8E,2023ApJ...944L..55L}, with NIRCam imaging in the F150W, F200W, F356W, and F444W filters and MIRI imaging in the F560W, F770W, and F1500W filters. \footnote{The JWST data presented in this paper were obtained from the Mikulski Archive for Space Telescopes (MAST) at the Space Telescope Science Institute. The specific observations analyzed can be accessed via \dataset[DOI: 10.17909/g4k8-3922]{https://doi.org/10.17909/g4k8-3922}.} We obtain images from the MAST archive and align the astrometry to GAIA DR3 \citep{2016A&A...595A...1G,2023A&A...674A...1G}. To remove the foreground emission from VV114, we first apply standard background subtraction, then identify point sources with a flux ratio $\log(S_\mathrm{F356W}/S_\mathrm{F200W})<-0.05$ as bright stars in VV114 and subtract them using the point spread function (PSF) model built with the \texttt{WebbPSF}\footnote{\url{https://github.com/spacetelescope/webbpsf}} package before performing photometry.
\section{Analysis and Results}\label{sec:analysis}
\subsection{JWST and ALMA Look at J0107a}\label{subsec:morphology}
\begin{figure*}
    \plotone{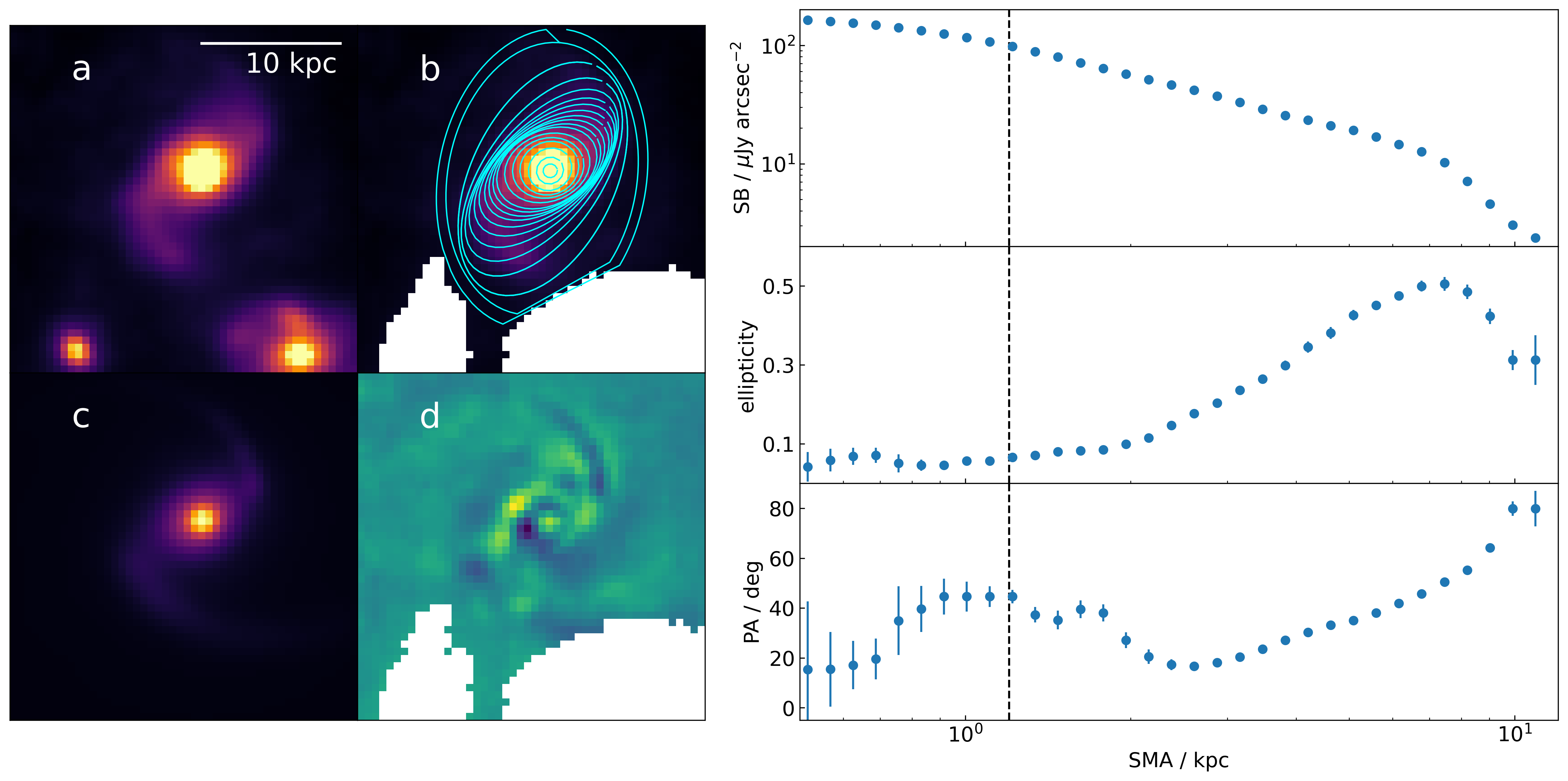}
    \caption{Left: a) $3\arcsec\times3\arcsec$ F444W cutout of J0107a displayed in $-2\textendash80\sigma$ range; b) the ellipse isophote fit to J0107a (cyan); c) the best-fit 2D parametric model of J0107a; d) the residual of the best-fit model. During modeling processes, the two neighboring galaxies are masked. Right: radial surface brightness (SB, top), ellipticity (middle), and position angle (PA, bottom) distributions of J0107a. The vertical dashed line marks the spatial resolution in the F444W band.}
    \label{fig:fig2}
\end{figure*}

The left panel of Figure \ref{fig:fig1}b shows the NIRCam image of J0107a. J0107a is clearly resolved, showing a grand-design barred spiral structure with two arms and a blue central point-like source, in line with an active galactic nucleus (AGN) as indicated by its X-ray detection in Chandra ACIS-S data \citep{2006ApJ...648..310G}. The bar structure is also visible in the ALMA band 7 continuum image with $0\farcs18\times0\farcs15$ beam size, implying star formation in the bar (Figure \ref{fig:fig1}b, the middle panel). To quantify the bar structure, we fit ellipse isophotes to the F444W image (Figure \ref{fig:fig2}). The projected bar size is determined as the semimajor axis (SMA) at maximum ellipticity $e_\mathrm{max}$. For J0107a's stellar bar, we find an SMA of 7.5 kpc and $e_\mathrm{max}=0.50$. The $e_\mathrm{max}$ is similar to that measured by \citet{2023ApJ...945L..10G} for $z>1$ barred galaxies in the CEERS survey, but the SMA is longer. We also perform parametric modeling of the F444W image, assuming a PSF model for the AGN, three S\'{e}rsic profiles for the bar, bulge, and disk, and two logarithmic curves for the arms \citep{1981AJ.....86.1847K}. We implement the models using the \texttt{numpy}, \texttt{scipy}, and \texttt{astropy} packages, then use the \texttt{dynesty} code \citep{2020MNRAS.493.3132S} to sample the joint posterior probability density function (PDF) of model parameters. The best-fit model and residual are shown in panels c and d of Figure \ref{fig:fig2}, respectively.  The rest-frame NIR light is dominated by the three S\'{e}rsic components with light fraction of $11.8_{-3.0}^{+2.3}\%$, $80.5_{-5.0}^{+4.6}\%$ and $7.9_{-3.5}^{+4.3}\%$ for the AGN, sum of S\'{e}rsic components, and spiral arms, respectively.
\par
J0107a is detected with three different molecular gas tracers, namely CO, [CI], and dust continuum. The derived $M_\mathrm{mol}$ ranges from $(9.2\pm1.7)\times10^{10}M_\odot$ using the CO(3-2) line to $(3.2\pm1.6)\times10^{11}M_\odot$ using the dust continuum,  assuming SMG/ULIRG-like CO excitation and conversion factors \citep{2021ApJ...917...94M}. If normal star-forming galaxies-like conversions \citep{2010ApJ...713..686D,2015A&A...577A..46D,2016ApJ...833...70D} are adopted, the $M_\mathrm{mol}$ will be $\sim6\times10^{11}M_\odot$. Therefore, J0107a is a very gas-rich DSFG with gas fraction $\sim20\textendash60\%$. No companion of J0107a is found in the band 3 cube. Assuming $\alpha_\mathrm{CO}=3.6M_\odot$/(K km s$^{-1}$ pc$^2$), $r_{31}=0.42\pm0.07$ and a Gaussian line profile with FWHM of 300 km s$^{-1}$, the $3\sigma$ sensitivity of 1.45 mJy beam$^{-1}$ in 50 km s$^{-1}$ channels translates to an upper limit of $M_\mathrm{mol}\lesssim10^{11}M_\odot$, implying the absence of any companion with molecular gas mass comparable with that of J0107a.
\subsection{Photometry and Spectral Energy Distribution
Modeling}\label{subsec:sedfitting}
\begin{figure*}
    \plotone{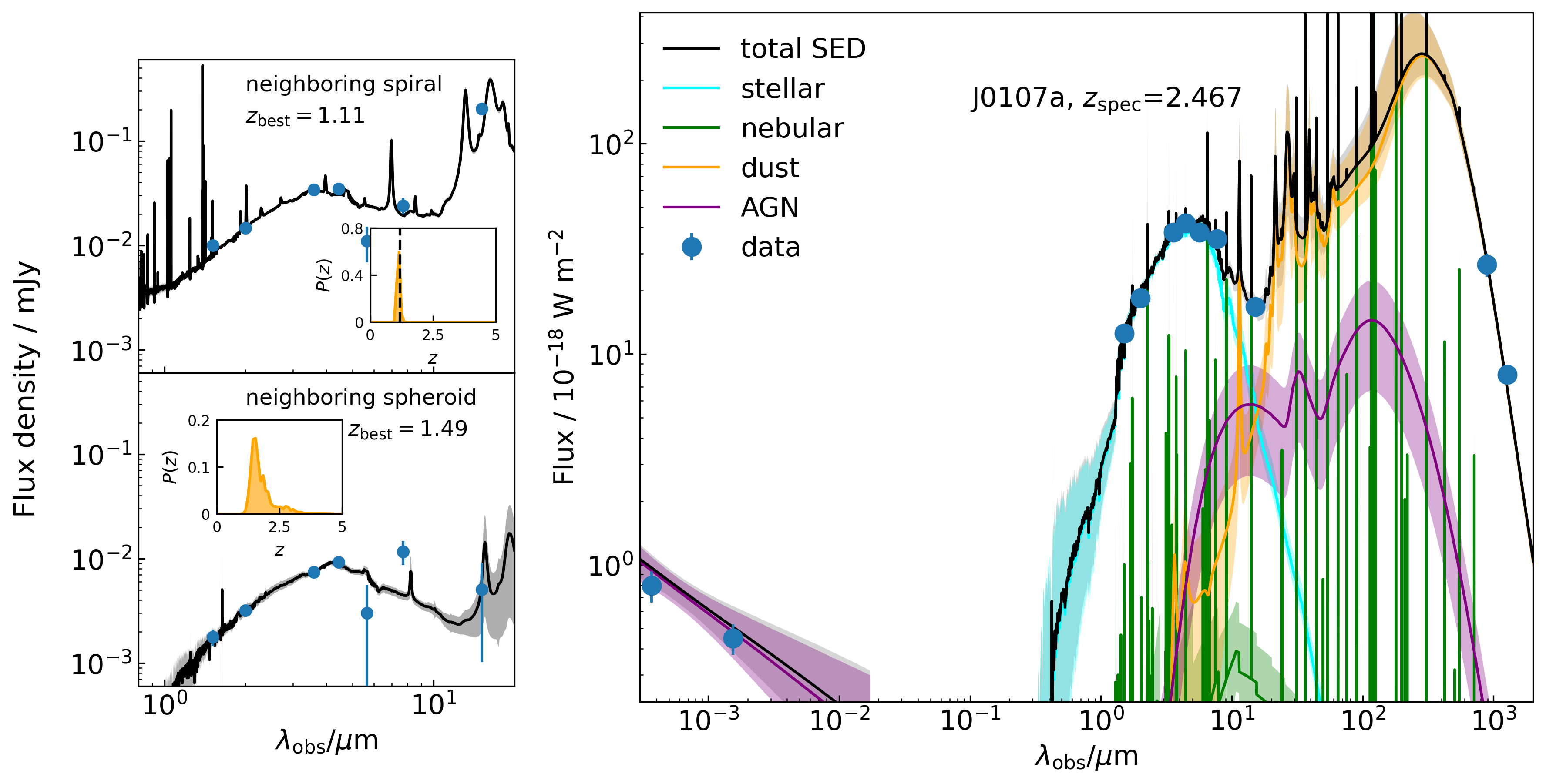}
    \caption{Left: \texttt{EAZY} fit of the JWST photometry (black circles and error bars). The median and 16th-84th percentile ranges of the SED at the best-fit redshift $z_\mathrm{best}$ are shown with the solid lines and shaded regions, respectively. The insert panels show the PDFs of photometric redshift. The photometric redshifts discussed in Section $\ref{subsec:neighbors}$ are 16th-50th-84th percentile values of the PDF. The vertical dashed lines mark the spectroscopic redshift identification.  Right: X-ray-to-millimeter SED fit of J0107a at $z_\mathrm{spec}=2.467$. For the total SED model and individual components, we show median and 16th-84th percentile ranges with the solid lines and shaded regions, respectively.}
    \label{fig:fig3}
\end{figure*}
To derive the photometry in JWST bands, we first combine the F356W and F444W images and then run source detection routines in the \texttt{photutils} package to generate Kron apertures for J0107a and the two neighboring galaxies in the combined $5\arcsec\times5\arcsec$ cutout image. The fluxes are measured with the Kron apertures and corrected for aperture loss using the PSF in each band. Flux uncertainties are estimated from sky by taking the standard deviations of fluxes in 1000 apertures randomly placed at blank positions within $5\arcsec$ radius from J0107a.
\par
We fit the observed spectral energy distribution (SED) of J0107a using the \texttt{CIGALE} code \citep{2019A&A...622A.103B}. In addition to the JWST data, we include ALMA band 6/7 continuum and Chandra X-ray fluxes taken from \citet{2014ApJ...781L..39T} and \citet{2021ApJ...917...94M}. We assume a nonparametric star formation history \citep{2019ApJ...877..140L} and adopt \citet{2009A&A...507.1793N} dust attenuation laws. The AGN emission is modeled with a power-law X-ray component plus the SKIRTOR library \citep{2016MNRAS.458.2288S}. For dust emission, we use the \citet{2014ApJ...780..172D} templates. We fix the redshift of J0107a at $z_\mathrm{spec}=2.467$ in \texttt{CIGALE} SED fitting.
\par
The SED fit of J0107a is shown in the right panel of Figure \ref{fig:fig3}. From SED fitting we derive the basic physical parameters of J0107a as follows:  $M_\star=4.5_{-1.4}^{+1.7}\times10^{11}M_\odot$, SFR=$499_{-179}^{+357}M_\odot$ yr$^{-1}$, dust mass $M_\mathrm{dust}=4.3_{-1.0}^{+1.0}\times10^{9}M_\odot$, and rest-frame 2-10 keV luminosity of $4.9_{-0.9}^{+0.9}\times10^{43}$ erg s$^{-1}$. The SFR of J0107a is $0.8_{-0.4}^{+0.9}$ times of the main sequence value of star-forming galaxies with the same redshift and $M_\star$ \citep{2014ApJS..214...15S}.
\subsection{Properties of the Two Neighboring galaxies and Association with J0107a}\label{subsec:neighbors}
\par
Two neighboring galaxies are visible within $2\farcs5$ from J0107a. The spiral galaxy  $1\farcs7$ southwest of J0107a exhibits a single-peaked redshift PDF (Figure \ref{fig:fig3}, the top-left panel). This PDF, derived from the SED fit of the JWST photometry using the \texttt{EAZY}\footnote{\url{https://github.com/gbrammer/eazy-py}} \citep{2008ApJ...686.1503B} code, yields a photometric redshift of $z_\mathrm{phot}=1.12_{-0.03}^{+0.05}$. An emission line at the central frequency of 105.458 GHz is detected at the position of the southwest spiral galaxy in the ALMA band 3 spectral cube (Figure \ref{fig:fig1}b, the middle panel), with the direction of velocity gradient aligning with the NIR major axis. Therefore, we are able to identify the emission line as redshifted CO(2-1) and derive spectroscopic redshift $z_\mathrm{spec}=1.186$ for this galaxy. The neighboring spheroid galaxy $1\farcs6$ southeast to J0107a has $z_\mathrm{phot}=1.69_{-0.26}^{+0.75}$ from the redshift PDF (Figure \ref{fig:fig3}, the bottom-left panel) and no line or continuum detection in the ALMA data, so its relation with J0107a remains uncertain.  If this spheroid galaxy is at the same redshift of J0107a, the $M_\star$ and SFR derived from SED fit of the JWST data will be $6.9_{-1.3}^{+1.2}\times10^{10}M_\odot$ and $4_{-4}^{+18}M_\odot$ yr$^{-1}$, respectively, and the stellar mass ratio of 1:6 does not satisfy the definition of a major merger.
\subsection{Dynamical Modeling}\label{subsec:diskfitting}
\begin{figure}
    \plotone{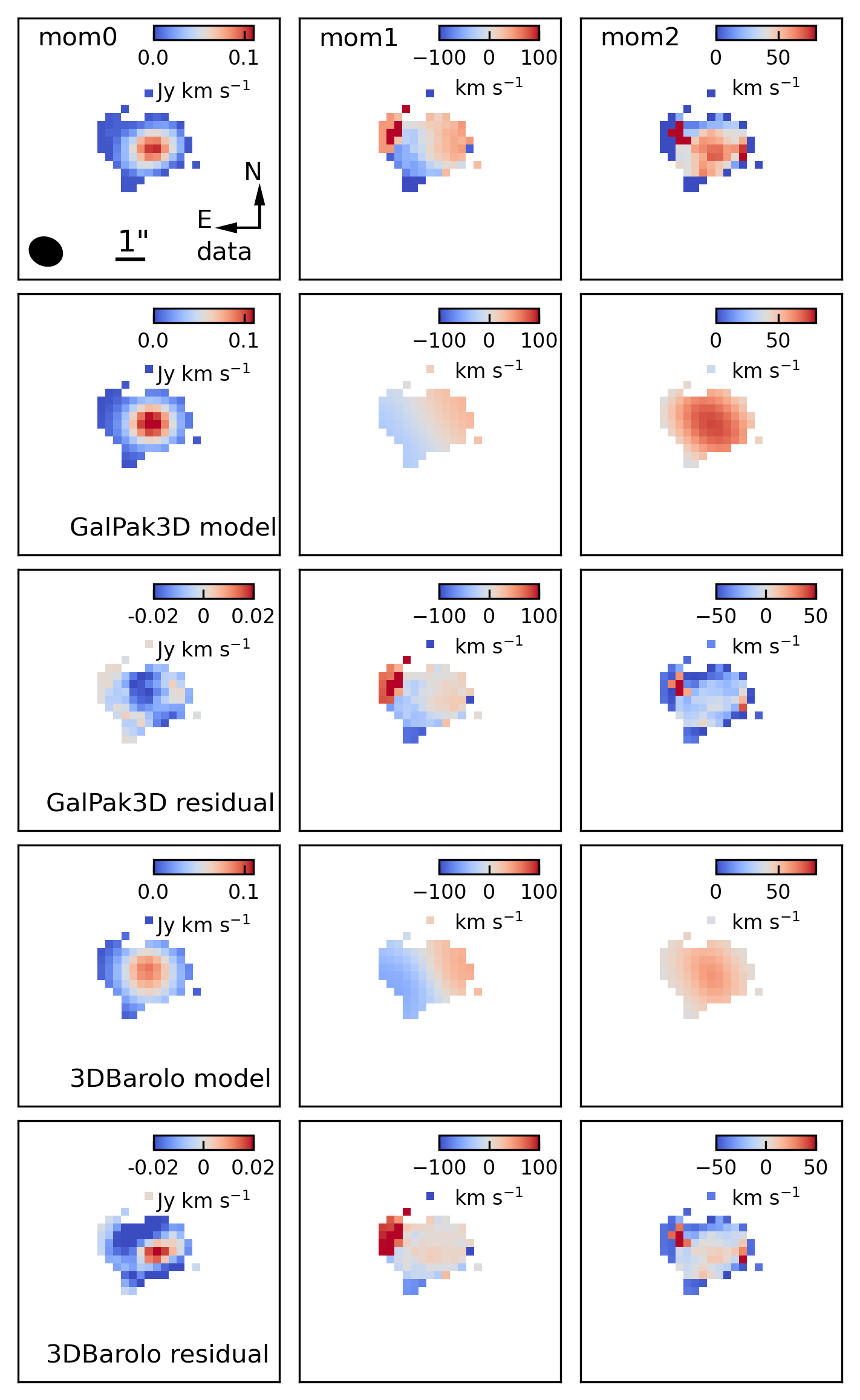}
    \caption{The zeroth, first, and second moment maps of the CO(4-3) emission line of J0107a. Also shown are the \texttt{GalPak3D}/\texttt{3D-Barolo} models and corresponding residual maps. The black ellipse in the top left panel indicates the beam size. The residual in the first moment map with redshifted velocity on the northeast edge overlaps with a horn-like structure in the zeroth moment map (Figure \ref{fig:fig1}b, the right panel) and may originate from gas inflow. However, the S/N is low so further observation is needed to confirm it.}
    \label{fig:fig4}
\end{figure}

A clear velocity gradient is seen in the velocity fields of all three spectral lines (Figure \ref{fig:fig4}, see also \citet{2021ApJ...917...94M}). We choose CO(4-3) to perform kinematic modeling because it has the highest S/N. We create a $9\farcs3\times9\farcs3$ cutout cube centered at J0107a with $20$ km s$^{-1}$ channel width and model the data with the \texttt{GalPak3D} code \citep{2015AJ....150...92B}, assuming an exponential disk profile and arctan rotation curve. This gives an inclination of $9\fdg7\pm0\fdg8$, half-light radius $r_\mathrm{e}=0\farcs54\pm0\farcs01$, maximum rotation velocity $v_\mathrm{MAX}=281\pm39$ km s$^{-1}$, and intrinsic velocity dispersion $\sigma=35\pm3$ km s$^{-1}$. In addition, we fit the data with the \texttt{3D-Barolo} code \citep{2015MNRAS.451.3021D} and find consistent values of an inclination of $10\fdg4$, maximum rotation velocity of $283$ km s$^{-1}$ and intrinsic velocity dispersion of 31 km s$^{-1}$. The best-fit models and residual maps are shown in Figure \ref{fig:fig4}. These results show J0107a has a dynamically cold gas disk with $v_\mathrm{MAX}/\sigma\sim8$.
\par
The dynamical mass within a radius of $2.2r_{e}=9.8$ kpc at $z=2.467$ can be calculated as:
\begin{equation}
    M_\mathrm{dyn}=\frac{2.2r_\mathrm{e}v^{2}_{\mathrm{circ}}}{G}
\end{equation}
where $G$ is the gravitational constant and $v_{\mathrm{circ}}=\sqrt{v_\mathrm{MAX}^{2}+4.4\sigma^{2}}$ is the circularized velocity \citep{2010ApJ...725.2324B} computed using the results from \texttt{GalPak3D} modelling. The resulting $M_\mathrm{dyn}$ of $\sim2\times10^{11}M_\odot$ is  less than the sum of $M_\star$ and $M_\mathrm{mol}$, indicating the total mass within a radius of 10 kpc is dominated by baryon.
\par
Previous analysis in \citet{2021ApJ...917...94M} reported a dynamical mass of $\sim10^{10}M_\odot$ based on similar velocity dispersion but a much lower $v_\mathrm{MAX}=69$ km s$^{-1}$, and their $M_\mathrm{dyn}$ conflicts with $M_\star$ and $M_\mathrm{mol}$, mainly because of their initial guess of inclination of $60\arcdeg$. We have tested initial guesses of $10\arcdeg$, $30\arcdeg$, and $60\arcdeg$, and let the value vary from $0\arcdeg$ to $90\arcdeg$. For initial guesses of $10\arcdeg$, $30\arcdeg$, the fit converges to the same results presented above, while for the initial guess of $60\arcdeg$, the fit converges to $\sim60\arcdeg$. Apparently, the inclination cannot be constrained alone with the $1\farcs2$ resolution CO data even the peak S/N reaches $\sim50$, and the new low inclination solution seems more appropriate. The outermost isophote of J0107a has an ellipticity of $\sim0.3$ which translates to an inclination of $<45\arcdeg$. The inclination from isophote fitting should be treated as an upper limit because of distortion of the isophotes by strong spiral arms \citep{2020ApJ...900..150Y} and  insufficient depth of the F444W image to detect the underlying smooth stellar disk. We caution that the current $M_\mathrm{dyn}$ estimate can still have large uncertainty due to the low inclination, beam smearing effect, and a possible more complex rotation pattern due to the bar. Future high-resolution CO observations are needed to better constrain the dynamics and mass distribution.

\section{Discussion and Conclusion}\label{sec:summary}
 A bright DSFG with a flux density of $\sim8$ mJy at the observed wavelength of 888 $\mu$m is expected to be the product of a gas-rich major merger according to the conventional wisdom. Nonetheless, J0107a's symmetric barred spiral morphology, lack of massive and gas-rich companion, and large dynamically cold gas disk all support a non-merger origin. Instead of merger-induced starburst, a more plausible formation channel would be cold mode accretion via inflowing gas streams from the cosmic web which penetrate the hot medium in the host halo and reach the central galaxy to fuel star formation and AGN activity \cite[e.g.,][]{2009MNRAS.395..160K}. Assuming the stellar-to-halo mass ratio $\log(M_\star/M_\mathrm{halo})$ ranges from $-1.5$ dex to $-1.8$ dex \citep{2021MNRAS.504..172S}, J0107a has a halo mass of $M_\mathrm{halo}=(1\textendash3)\times10^{13}M_\odot$, which lies at the boundary between hot and cold in hot modes of accretion \citep{2006MNRAS.368....2D} and represents the most efficient normal star-forming galaxy in converting gas into stars at this redshift.
 The predicted baryon accretion rate is $1500\textendash5400M_\odot$ yr$^{-1}$ \citep{2009Natur.457..451D}, which is sufficient to supply J0107a's star formation. If J0107a is indeed fed by such streams, it will be an extreme case with the maximum allowed accretion rate, thus an ideal target to directly detect such streams onto a bright DSFG and demonstrate their heterogeneous origins.
\par
The large stellar bar of J0107a is of particular interest in the context of galaxy assembly. J0107a has the highest redshift among known $z>1$ galaxies with a clear stellar bar identified \citep[e.g.,][]{2014MNRAS.445.3466S,2023ApJ...945L..10G}, but also see a lensed case at $z=4.26$ \citep{2023arXiv230616039S} and potential gas/dust bars at $z=1.5\textendash4.4$ \citep{2019ApJ...876..130H,2023arXiv230814798T}.  In local strongly barred galaxies, the global SFR is suppressed compared to unbarred spiral galaxies with similar mass \citep[e.g.,][]{2015A&A...580A.116G,2021MNRAS.507.4389G}, while J0107a lies above the $z\sim2.5$ star-forming main sequence. This may be interpreted as the bar-accelerated stellar mass build-up in the past \citep[e.g.,][]{2017ApJ...845...93K,2020MNRAS.499.1116F} due to the effective gas redistribution as we will discuss below, and J0107a is caught in the early starburst phase of a massive barred galaxy.
\par
 In addition to the stellar bar, the 888 $\mu$m dust continuum map suggests the presence of a dust/gas bar (Figure \ref{fig:fig1}b, the middle panel), which is the site of star formation. The dust emission shows three peaks and is reminiscent of the enhancement of SFR at the galactic center and the bar ends in local barred galaxies \citep{2023ApJ...943....7M}. Elevated central star formation is common in local barred galaxies \citep[e.g.,][]{2019MNRAS.484.5192C,2020MNRAS.499.1406L} and explained by theoretical studies as a consequence of gas being funneled by the bar toward the center and feeds the star formation and supermassive black hole \citep[e.g.,][]{2016MNRAS.463.1074C}. The higher central molecular gas concentration observed in local barred spiral galaxies \citep{2007PASJ...59..117K,2022A&A...666A.175Y} provides direct evidence for bar-driven gas inflow, and \citet{2022A&A...666A.175Y} also suggest the spiral arms  can playing a role in transporting gas to the center. On the other hand, barred galaxies with suppressed central SFR often exhibit weak to nearly vanished spiral arms \citep{2020ApJ...893...19W,2022A&A...661A..98Y}. J0107a possesses a strong stellar bar and pronounced spiral arms, and the high SFR may be triggered by gas inflow driven by stellar structures in its gas-rich disk.
\par
Numerical simulations have shown that an isolated baryon-dominated and dynamically cold disk allows bar formation at early times \citep{2022MNRAS.512.5339R,2023ApJ...947...80B}, though it has also been suggested that a high gas fraction can weaken the bar and slows down its formation \citep[e.g.,][]{2013MNRAS.429.1949A}. Alternative formation channels of barred spiral structures include interaction with a companion galaxy \citep[e.g.,][]{1990ApJ...364..415E,2008ApJ...683...94O}. The photometric redshift of the southeast spheroid galaxy does not exclude the possibility of it being a companion of J0107a. From the analysis of the JWST data, the spheroid galaxy will have $M_\star$ of $\sim1/6\times$ J0107a and a distance of 13 kpc to J0107a if it is at $z=2.467$, which is sufficient to affect J0107a's disk. While the southeast spiral arm indeed seems brighter, tidal features were not found.  Furthermore, J0107a's CO disk is dynamically cold and dominated by ordered rotation ($V_\mathrm{max}/\sigma\sim8$). Considering that cold accretion can also bring a low $V_\mathrm{max}/\sigma$ component  \citep{2007ApJ...668...94H}, secular bar formation is thus favored and consistent with the first massive and well-ordered disks emerging in the cosmic noon \citep{2017ApJ...843...46S}.
\par
Previous simulations of bar formation have neither covered the parameter space with $M_\star$, SFR, and gas fraction as high as J0107a, nor gas accretion from the surrounding environment. Refined simulations reproducing the properties of J0107a are needed to understand the formation mechanisms of stellar and gas bars in the earliest massive disks. In terms of observations, deep high-resolution spectroscopy is needed to confirm the isolated disk nature of J0107a and to investigate the interplay between bar, gas, and star formation.
\par
Finally, J0107a has been mis-identified as a strongly lensed SMG magnified by approximately ten times in \citet{2021ApJ...917...94M} to explain the bright CO(4-3) emission with narrow line FWHM (FWHM = $193\pm11$ km s$^{-1}$; $S\Delta v=3.0\pm0.2$ Jy km s$^{-1}$). However, it is now clear that this is caused by a low inclination of $\sim10\arcdeg$, as the regular morphology and absence of a nearby massive source can rule out strong gravitational lensing effect. While the empirical relation between CO line FWHM and velocity-integrated intensity \citep[e.g.,][]{2012ApJ...752..152H,2013MNRAS.429.3047B} is a natural consequence of the virial theorem for large samples of unlensed galaxies, the use of such relations to identify individual lensed source should be careful as the result is sensitive to the unknown inclination.

\begin{acknowledgements}
    We are grateful to the anonymous reviewer for the helpful and constructive comments that improved the quality of this manuscript.
    SM is supported by Japan Society for the Promotion of Science (JSPS) KAKENHI with the Grant number of 23KJ0450.  YT is supported by JSPS KAKENHI (No. 22H04939).
    This paper makes use of the following ALMA data: ADS/JAO.ALMA \#2013.1.00469.S,  \#2013.1.01057.S, \#2015.1.00973.S, \#2015.1.00902.S, and \#2013.1.00740.S. ALMA is a partnership of ESO (representing its member states), NSF (USA), and NINS (Japan), together with NRC (Canada), MOST and ASIAA (Taiwan), and KASI (Republic of Korea), in cooperation with the Republic of Chile. The Joint ALMA Observatory is operated by ESO, AUI/NRAO and NAOJ. The Flatiron Institute is supported by the Simons Foundation.
\end{acknowledgements}

\software{numpy \citep{harris2020array},
scipy \citep{2020SciPy-NMeth},
matplotlib \citep{Hunter:2007},
astropy \citep{2018AJ....156..123A}, photutils \citep{2022zndo...6825092B}, dynesty \citep{2020MNRAS.493.3132S}}

\bibliographystyle{aasjournal}
\bibliography{main}

\begin{thebibliography}{}
\expandafter\ifx\csname natexlab\endcsname\relax\def\natexlab#1{#1}\fi
\providecommand{\url}[1]{\href{#1}{#1}}
\providecommand{\dodoi}[1]{doi:~\href{http://doi.org/#1}{\nolinkurl{#1}}}
\providecommand{\doeprint}[1]{\href{http://ascl.net/#1}{\nolinkurl{http://ascl.net/#1}}}
\providecommand{\doarXiv}[1]{\href{https://arxiv.org/abs/#1}{\nolinkurl{https://arxiv.org/abs/#1}}}

\bibitem[{{Astropy Collaboration} {et~al.}(2018){Astropy Collaboration},
  {Price-Whelan}, {Sip{\H{o}}cz}, {G{\"u}nther}, {Lim}, {Crawford}, {Conseil},
  {Shupe}, {Craig}, {Dencheva}, {Ginsburg}, {VanderPlas}, {Bradley},
  {P{\'e}rez-Su{\'a}rez}, {de Val-Borro}, {Aldcroft}, {Cruz}, {Robitaille},
  {Tollerud}, {Ardelean}, {Babej}, {Bach}, {Bachetti}, {Bakanov}, {Bamford},
  {Barentsen}, {Barmby}, {Baumbach}, {Berry}, {Biscani}, {Boquien}, {Bostroem},
  {Bouma}, {Brammer}, {Bray}, {Breytenbach}, {Buddelmeijer}, {Burke},
  {Calderone}, {Cano Rodr{\'\i}guez}, {Cara}, {Cardoso}, {Cheedella}, {Copin},
  {Corrales}, {Crichton}, {D'Avella}, {Deil}, {Depagne}, {Dietrich}, {Donath},
  {Droettboom}, {Earl}, {Erben}, {Fabbro}, {Ferreira}, {Finethy}, {Fox},
  {Garrison}, {Gibbons}, {Goldstein}, {Gommers}, {Greco}, {Greenfield},
  {Groener}, {Grollier}, {Hagen}, {Hirst}, {Homeier}, {Horton}, {Hosseinzadeh},
  {Hu}, {Hunkeler}, {Ivezi{\'c}}, {Jain}, {Jenness}, {Kanarek}, {Kendrew},
  {Kern}, {Kerzendorf}, {Khvalko}, {King}, {Kirkby}, {Kulkarni}, {Kumar},
  {Lee}, {Lenz}, {Littlefair}, {Ma}, {Macleod}, {Mastropietro}, {McCully},
  {Montagnac}, {Morris}, {Mueller}, {Mumford}, {Muna}, {Murphy}, {Nelson},
  {Nguyen}, {Ninan}, {N{\"o}the}, {Ogaz}, {Oh}, {Parejko}, {Parley}, {Pascual},
  {Patil}, {Patil}, {Plunkett}, {Prochaska}, {Rastogi}, {Reddy Janga},
  {Sabater}, {Sakurikar}, {Seifert}, {Sherbert}, {Sherwood-Taylor}, {Shih},
  {Sick}, {Silbiger}, {Singanamalla}, {Singer}, {Sladen}, {Sooley},
  {Sornarajah}, {Streicher}, {Teuben}, {Thomas}, {Tremblay}, {Turner},
  {Terr{\'o}n}, {van Kerkwijk}, {de la Vega}, {Watkins}, {Weaver}, {Whitmore},
  {Woillez}, {Zabalza}, \& {Astropy Contributors}}]{2018AJ....156..123A}
{Astropy Collaboration}, {Price-Whelan}, A.~M., {Sip{\H{o}}cz}, B.~M., {et~al.}
  2018, \aj, 156, 123, \dodoi{10.3847/1538-3881/aabc4f}

\bibitem[{{Athanassoula} {et~al.}(2013){Athanassoula}, {Machado}, \&
  {Rodionov}}]{2013MNRAS.429.1949A}
{Athanassoula}, E., {Machado}, R. E.~G., \& {Rodionov}, S.~A. 2013, \mnras,
  429, 1949, \dodoi{10.1093/mnras/sts452}

\bibitem[{{Bland-Hawthorn} {et~al.}(2023){Bland-Hawthorn}, {Tepper-Garcia},
  {Agertz}, \& {Freeman}}]{2023ApJ...947...80B}
{Bland-Hawthorn}, J., {Tepper-Garcia}, T., {Agertz}, O., \& {Freeman}, K. 2023,
  \apj, 947, 80, \dodoi{10.3847/1538-4357/acc469}

\bibitem[{{Boquien} {et~al.}(2019){Boquien}, {Burgarella}, {Roehlly}, {Buat},
  {Ciesla}, {Corre}, {Inoue}, \& {Salas}}]{2019A&A...622A.103B}
{Boquien}, M., {Burgarella}, D., {Roehlly}, Y., {et~al.} 2019, \aap, 622, A103,
  \dodoi{10.1051/0004-6361/201834156}

\bibitem[{{Bothwell} {et~al.}(2013){Bothwell}, {Smail}, {Chapman}, {Genzel},
  {Ivison}, {Tacconi}, {Alaghband-Zadeh}, {Bertoldi}, {Blain}, {Casey}, {Cox},
  {Greve}, {Lutz}, {Neri}, {Omont}, \& {Swinbank}}]{2013MNRAS.429.3047B}
{Bothwell}, M.~S., {Smail}, I., {Chapman}, S.~C., {et~al.} 2013, \mnras, 429,
  3047, \dodoi{10.1093/mnras/sts562}

\bibitem[{{Bouch{\'e}} {et~al.}(2015){Bouch{\'e}}, {Carfantan}, {Schroetter},
  {Michel-Dansac}, \& {Contini}}]{2015AJ....150...92B}
{Bouch{\'e}}, N., {Carfantan}, H., {Schroetter}, I., {Michel-Dansac}, L., \&
  {Contini}, T. 2015, \aj, 150, 92, \dodoi{10.1088/0004-6256/150/3/92}

\bibitem[{{Bradley} {et~al.}(2022){Bradley}, {Sip{\H{o}}cz}, {Robitaille},
  {Tollerud}, {Vin{\'\i}cius}, {Deil}, {Barbary}, {Wilson}, {Busko}, {Donath},
  {G{\"u}nther}, {Cara}, {Lim}, {Me{\ss}linger}, {Conseil}, {Bostroem},
  {Droettboom}, {Bray}, {Andersen Bratholm}, {Barentsen}, {Craig}, {Rathi},
  {Pascual}, {Perren}, {Georgiev}, {De Val-Borro}, {Kerzendorf}, {Bach},
  {Quint}, \& {Souchereau}}]{2022zndo...6825092B}
{Bradley}, L., {Sip{\H{o}}cz}, B., {Robitaille}, T., {et~al.} 2022,
  {astropy/photutils: 1.5.0}, 1.5.0, Zenodo,  Zenodo,
  \dodoi{10.5281/zenodo.6825092}

\bibitem[{{Brammer} {et~al.}(2008){Brammer}, {van Dokkum}, \&
  {Coppi}}]{2008ApJ...686.1503B}
{Brammer}, G.~B., {van Dokkum}, P.~G., \& {Coppi}, P. 2008, \apj, 686, 1503,
  \dodoi{10.1086/591786}

\bibitem[{{Burkert} {et~al.}(2010){Burkert}, {Genzel}, {Bouch{\'e}}, {Cresci},
  {Khochfar}, {Sommer-Larsen}, {Sternberg}, {Naab}, {F{\"o}rster Schreiber},
  {Tacconi}, {Shapiro}, {Hicks}, {Lutz}, {Davies}, {Buschkamp}, \&
  {Genel}}]{2010ApJ...725.2324B}
{Burkert}, A., {Genzel}, R., {Bouch{\'e}}, N., {et~al.} 2010, \apj, 725, 2324,
  \dodoi{10.1088/0004-637X/725/2/2324}

\bibitem[{{Carles} {et~al.}(2016){Carles}, {Martel}, {Ellison}, \&
  {Kawata}}]{2016MNRAS.463.1074C}
{Carles}, C., {Martel}, H., {Ellison}, S.~L., \& {Kawata}, D. 2016, \mnras,
  463, 1074, \dodoi{10.1093/mnras/stw2056}

\bibitem[{{CASA Team} {et~al.}(2022){CASA Team}, {Bean}, {Bhatnagar}, {Castro},
  {Donovan Meyer}, {Emonts}, {Garcia}, {Garwood}, {Golap}, {Gonzalez Villalba},
  {Harris}, {Hayashi}, {Hoskins}, {Hsieh}, {Jagannathan}, {Kawasaki},
  {Keimpema}, {Kettenis}, {Lopez}, {Marvil}, {Masters}, {McNichols},
  {Mehringer}, {Miel}, {Moellenbrock}, {Montesino}, {Nakazato}, {Ott}, {Petry},
  {Pokorny}, {Raba}, {Rau}, {Schiebel}, {Schweighart}, {Sekhar}, {Shimada},
  {Small}, {Steeb}, {Sugimoto}, {Suoranta}, {Tsutsumi}, {van Bemmel},
  {Verkouter}, {Wells}, {Xiong}, {Szomoru}, {Griffith}, {Glendenning}, \&
  {Kern}}]{2022PASP..134k4501C}
{CASA Team}, {Bean}, B., {Bhatnagar}, S., {et~al.} 2022, \pasp, 134, 114501,
  \dodoi{10.1088/1538-3873/ac9642}

\bibitem[{{Casey} {et~al.}(2014){Casey}, {Narayanan}, \&
  {Cooray}}]{2014PhR...541...45C}
{Casey}, C.~M., {Narayanan}, D., \& {Cooray}, A. 2014, \physrep, 541, 45,
  \dodoi{10.1016/j.physrep.2014.02.009}

\bibitem[{{Chabrier}(2003)}]{2003PASP..115..763C}
{Chabrier}, G. 2003, \pasp, 115, 763, \dodoi{10.1086/376392}

\bibitem[{{Chen} {et~al.}(2015){Chen}, {Smail}, {Swinbank}, {Simpson}, {Ma},
  {Alexander}, {Biggs}, {Brandt}, {Chapman}, {Coppin}, {Danielson},
  {Dannerbauer}, {Edge}, {Greve}, {Ivison}, {Karim}, {Menten}, {Schinnerer},
  {Walter}, {Wardlow}, {Wei{\ss}}, \& {van der Werf}}]{2015ApJ...799..194C}
{Chen}, C.-C., {Smail}, I., {Swinbank}, A.~M., {et~al.} 2015, \apj, 799, 194,
  \dodoi{10.1088/0004-637X/799/2/194}

\bibitem[{{Chen} {et~al.}(2022{\natexlab{a}}){Chen}, {Liao}, {Smail},
  {Swinbank}, {Ao}, {Bunker}, {Chapman}, {Hatsukade}, {Ivison}, {Lee},
  {Serjeant}, {Umehata}, {Wang}, \& {Zhao}}]{2022ApJ...929..159C}
{Chen}, C.-C., {Liao}, C.-L., {Smail}, I., {et~al.} 2022{\natexlab{a}}, \apj,
  929, 159, \dodoi{10.3847/1538-4357/ac61df}

\bibitem[{{Chen} {et~al.}(2022{\natexlab{b}}){Chen}, {Gao}, {Hsu}, {Liao},
  {Ling}, {Lo}, {Smail}, {Wang}, \& {Wang}}]{2022ApJ...939L...7C}
{Chen}, C.-C., {Gao}, Z.-K., {Hsu}, Q.-N., {et~al.} 2022{\natexlab{b}}, \apjl,
  939, L7, \dodoi{10.3847/2041-8213/ac98c6}

\bibitem[{{Cheng} {et~al.}(2023){Cheng}, {Huang}, {Smail}, {Yan}, {Cohen},
  {Jansen}, {Windhorst}, {Ma}, {Koekemoer}, {Willmer}, {Willner}, {Diego},
  {Frye}, {Conselice}, {Ferreira}, {Petric}, {Yun}, {Gim}, {Polletta},
  {Duncan}, {Holwerda}, {R{\"o}ttgering}, {Honor}, {Hathi}, {Kamieneski},
  {Adams}, {Coe}, {Broadhurst}, {Summers}, {Tompkins}, {Driver}, {Grogin},
  {Marshall}, {Pirzkal}, {Robotham}, \& {Ryan}}]{2023ApJ...942L..19C}
{Cheng}, C., {Huang}, J.-S., {Smail}, I., {et~al.} 2023, \apjl, 942, L19,
  \dodoi{10.3847/2041-8213/aca9d0}

\bibitem[{{Chown} {et~al.}(2019){Chown}, {Li}, {Athanassoula}, {Li}, {Wilson},
  {Lin}, {Mo}, {Parker}, \& {Xiao}}]{2019MNRAS.484.5192C}
{Chown}, R., {Li}, C., {Athanassoula}, E., {et~al.} 2019, \mnras, 484, 5192,
  \dodoi{10.1093/mnras/stz349}

\bibitem[{{da Cunha} {et~al.}(2015){da Cunha}, {Walter}, {Smail}, {Swinbank},
  {Simpson}, {Decarli}, {Hodge}, {Weiss}, {van der Werf}, {Bertoldi},
  {Chapman}, {Cox}, {Danielson}, {Dannerbauer}, {Greve}, {Ivison}, {Karim}, \&
  {Thomson}}]{2015ApJ...806..110D}
{da Cunha}, E., {Walter}, F., {Smail}, I.~R., {et~al.} 2015, \apj, 806, 110,
  \dodoi{10.1088/0004-637X/806/1/110}

\bibitem[{{Daddi} {et~al.}(2010){Daddi}, {Bournaud}, {Walter}, {Dannerbauer},
  {Carilli}, {Dickinson}, {Elbaz}, {Morrison}, {Riechers}, {Onodera}, {Salmi},
  {Krips}, \& {Stern}}]{2010ApJ...713..686D}
{Daddi}, E., {Bournaud}, F., {Walter}, F., {et~al.} 2010, \apj, 713, 686,
  \dodoi{10.1088/0004-637X/713/1/686}

\bibitem[{{Daddi} {et~al.}(2015){Daddi}, {Dannerbauer}, {Liu}, {Aravena},
  {Bournaud}, {Walter}, {Riechers}, {Magdis}, {Sargent}, {B{\'e}thermin},
  {Carilli}, {Cibinel}, {Dickinson}, {Elbaz}, {Gao}, {Gobat}, {Hodge}, \&
  {Krips}}]{2015A&A...577A..46D}
{Daddi}, E., {Dannerbauer}, H., {Liu}, D., {et~al.} 2015, \aap, 577, A46,
  \dodoi{10.1051/0004-6361/201425043}

\bibitem[{{Dav{\'e}} {et~al.}(2010){Dav{\'e}}, {Finlator}, {Oppenheimer},
  {Fardal}, {Katz}, {Kere{\v{s}}}, \& {Weinberg}}]{2010MNRAS.404.1355D}
{Dav{\'e}}, R., {Finlator}, K., {Oppenheimer}, B.~D., {et~al.} 2010, \mnras,
  404, 1355, \dodoi{10.1111/j.1365-2966.2010.16395.x}

\bibitem[{{Decarli} {et~al.}(2016){Decarli}, {Walter}, {Aravena}, {Carilli},
  {Bouwens}, {da Cunha}, {Daddi}, {Elbaz}, {Riechers}, {Smail}, {Swinbank},
  {Weiss}, {Bacon}, {Bauer}, {Bell}, {Bertoldi}, {Chapman}, {Colina}, {Cortes},
  {Cox}, {G{\'o}nzalez-L{\'o}pez}, {Inami}, {Ivison}, {Hodge}, {Karim},
  {Magnelli}, {Ota}, {Popping}, {Rix}, {Sargent}, {van der Wel}, \& {van der
  Werf}}]{2016ApJ...833...70D}
{Decarli}, R., {Walter}, F., {Aravena}, M., {et~al.} 2016, \apj, 833, 70,
  \dodoi{10.3847/1538-4357/833/1/70}

\bibitem[{{Dekel} \& {Birnboim}(2006)}]{2006MNRAS.368....2D}
{Dekel}, A., \& {Birnboim}, Y. 2006, \mnras, 368, 2,
  \dodoi{10.1111/j.1365-2966.2006.10145.x}

\bibitem[{{Dekel} {et~al.}(2009){Dekel}, {Birnboim}, {Engel}, {Freundlich},
  {Goerdt}, {Mumcuoglu}, {Neistein}, {Pichon}, {Teyssier}, \&
  {Zinger}}]{2009Natur.457..451D}
{Dekel}, A., {Birnboim}, Y., {Engel}, G., {et~al.} 2009, \nat, 457, 451,
  \dodoi{10.1038/nature07648}

\bibitem[{{Di Teodoro} \& {Fraternali}(2015)}]{2015MNRAS.451.3021D}
{Di Teodoro}, E.~M., \& {Fraternali}, F. 2015, \mnras, 451, 3021,
  \dodoi{10.1093/mnras/stv1213}

\bibitem[{{Draine} {et~al.}(2014){Draine}, {Aniano}, {Krause}, {Groves},
  {Sandstrom}, {Braun}, {Leroy}, {Klaas}, {Linz}, {Rix}, {Schinnerer},
  {Schmiedeke}, \& {Walter}}]{2014ApJ...780..172D}
{Draine}, B.~T., {Aniano}, G., {Krause}, O., {et~al.} 2014, \apj, 780, 172,
  \dodoi{10.1088/0004-637X/780/2/172}

\bibitem[{{Dudzevi{\v{c}}i{\={u}}t{\.{e}}}
  {et~al.}(2020){Dudzevi{\v{c}}i{\={u}}t{\.{e}}}, {Smail}, {Swinbank}, {Stach},
  {Almaini}, {da Cunha}, {An}, {Arumugam}, {Birkin}, {Blain}, {Chapman},
  {Chen}, {Conselice}, {Coppin}, {Dunlop}, {Farrah}, {Geach}, {Gullberg},
  {Hartley}, {Hodge}, {Ivison}, {Maltby}, {Scott}, {Simpson}, {Simpson},
  {Thomson}, {Walter}, {Wardlow}, {Weiss}, \& {van der
  Werf}}]{2020MNRAS.494.3828D}
{Dudzevi{\v{c}}i{\={u}}t{\.{e}}}, U., {Smail}, I., {Swinbank}, A.~M., {et~al.}
  2020, \mnras, 494, 3828, \dodoi{10.1093/mnras/staa769}

\bibitem[{{Elmegreen} {et~al.}(1990){Elmegreen}, {Elmegreen}, \&
  {Bellin}}]{1990ApJ...364..415E}
{Elmegreen}, D.~M., {Elmegreen}, B.~G., \& {Bellin}, A.~D. 1990, \apj, 364,
  415, \dodoi{10.1086/169424}

\bibitem[{{Engel} {et~al.}(2010){Engel}, {Tacconi}, {Davies}, {Neri}, {Smail},
  {Chapman}, {Genzel}, {Cox}, {Greve}, {Ivison}, {Blain}, {Bertoldi}, \&
  {Omont}}]{2010ApJ...724..233E}
{Engel}, H., {Tacconi}, L.~J., {Davies}, R.~I., {et~al.} 2010, \apj, 724, 233,
  \dodoi{10.1088/0004-637X/724/1/233}

\bibitem[{{Evans} {et~al.}(2022){Evans}, {Frayer}, {Charmandaris}, {Armus},
  {Inami}, {Surace}, {Linden}, {Soifer}, {Diaz-Santos}, {Larson}, {Rich},
  {Song}, {Barcos-Munoz}, {Mazzarella}, {Privon}, {U}, {Medling}, {B{\"o}ker},
  {Aalto}, {Iwasawa}, {Howell}, {van der Werf}, {Appleton}, {Bohn}, {Brown},
  {Hayward}, {Hoshioka}, {Kemper}, {Lai}, {Law}, {Malkan}, {Marshall},
  {Murphy}, {Sanders}, \& {Stierwalt}}]{2022ApJ...940L...8E}
{Evans}, A.~S., {Frayer}, D.~T., {Charmandaris}, V., {et~al.} 2022, \apjl, 940,
  L8, \dodoi{10.3847/2041-8213/ac9971}

\bibitem[{{Fraser-McKelvie} {et~al.}(2020){Fraser-McKelvie}, {Merrifield},
  {Arag{\'o}n-Salamanca}, {Peterken}, {Kraljic}, {Masters}, {Stark},
  {Fragkoudi}, {Smethurst}, {Boardman}, {Drory}, \&
  {Lane}}]{2020MNRAS.499.1116F}
{Fraser-McKelvie}, A., {Merrifield}, M., {Arag{\'o}n-Salamanca}, A., {et~al.}
  2020, \mnras, 499, 1116, \dodoi{10.1093/mnras/staa2866}

\bibitem[{{Gaia Collaboration} {et~al.}(2016){Gaia Collaboration}, {Prusti},
  {de Bruijne}, {Brown}, {Vallenari}, {Babusiaux}, {Bailer-Jones}, {Bastian},
  {Biermann}, {Evans}, {Eyer}, {Jansen}, {Jordi}, {Klioner}, {Lammers},
  {Lindegren}, {Luri}, {Mignard}, {Milligan}, {Panem}, {Poinsignon},
  {Pourbaix}, {Randich}, {Sarri}, {Sartoretti}, {Siddiqui}, {Soubiran},
  {Valette}, {van Leeuwen}, {Walton}, {Aerts}, {Arenou}, {Cropper}, {Drimmel},
  {H{\o}g}, {Katz}, {Lattanzi}, {O'Mullane}, {Grebel}, {Holland}, {Huc},
  {Passot}, {Bramante}, {Cacciari}, {Casta{\~n}eda}, {Chaoul}, {Cheek}, {De
  Angeli}, {Fabricius}, {Guerra}, {Hern{\'a}ndez}, {Jean-Antoine-Piccolo},
  {Masana}, {Messineo}, {Mowlavi}, {Nienartowicz}, {Ord{\'o}{\~n}ez-Blanco},
  {Panuzzo}, {Portell}, {Richards}, {Riello}, {Seabroke}, {Tanga},
  {Th{\'e}venin}, {Torra}, {Els}, {Gracia-Abril}, {Comoretto},
  {Garcia-Reinaldos}, {Lock}, {Mercier}, {Altmann}, {Andrae}, {Astraatmadja},
  {Bellas-Velidis}, {Benson}, {Berthier}, {Blomme}, {Busso}, {Carry},
  {Cellino}, {Clementini}, {Cowell}, {Creevey}, {Cuypers}, {Davidson}, {De
  Ridder}, {de Torres}, {Delchambre}, {Dell'Oro}, {Ducourant}, {Fr{\'e}mat},
  {Garc{\'\i}a-Torres}, {Gosset}, {Halbwachs}, {Hambly}, {Harrison}, {Hauser},
  {Hestroffer}, {Hodgkin}, {Huckle}, {Hutton}, {Jasniewicz}, {Jordan},
  {Kontizas}, {Korn}, {Lanzafame}, {Manteiga}, {Moitinho}, {Muinonen},
  {Osinde}, {Pancino}, {Pauwels}, {Petit}, {Recio-Blanco}, {Robin}, {Sarro},
  {Siopis}, {Smith}, {Smith}, {Sozzetti}, {Thuillot}, {van Reeven}, {Viala},
  {Abbas}, {Abreu Aramburu}, {Accart}, {Aguado}, {Allan}, {Allasia},
  {Altavilla}, {{\'A}lvarez}, {Alves}, {Anderson}, {Andrei}, {Anglada Varela},
  {Antiche}, {Antoja}, {Ant{\'o}n}, {Arcay}, {Atzei}, {Ayache}, {Bach},
  {Baker}, {Balaguer-N{\'u}{\~n}ez}, {Barache}, {Barata}, {Barbier}, {Barblan},
  {Baroni}, {Barrado y Navascu{\'e}s}, {Barros}, {Barstow}, {Becciani},
  {Bellazzini}, {Bellei}, {Bello Garc{\'\i}a}, {Belokurov}, {Bendjoya},
  {Berihuete}, {Bianchi}, {Bienaym{\'e}}, {Billebaud}, {Blagorodnova},
  {Blanco-Cuaresma}, {Boch}, {Bombrun}, {Borrachero}, {Bouquillon}, {Bourda},
  {Bouy}, {Bragaglia}, {Breddels}, {Brouillet}, {Br{\"u}semeister},
  {Bucciarelli}, {Budnik}, {Burgess}, {Burgon}, {Burlacu}, {Busonero}, {Buzzi},
  {Caffau}, {Cambras}, {Campbell}, {Cancelliere}, {Cantat-Gaudin}, {Carlucci},
  {Carrasco}, {Castellani}, {Charlot}, {Charnas}, {Charvet}, {Chassat},
  {Chiavassa}, {Clotet}, {Cocozza}, {Collins}, {Collins}, {Costigan}, {Crifo},
  {Cross}, {Crosta}, {Crowley}, {Dafonte}, {Damerdji}, {Dapergolas}, {David},
  {David}, {De Cat}, {de Felice}, {de Laverny}, {De Luise}, {De March}, {de
  Martino}, {de Souza}, {Debosscher}, {del Pozo}, {Delbo}, {Delgado},
  {Delgado}, {di Marco}, {Di Matteo}, {Diakite}, {Distefano}, {Dolding}, {Dos
  Anjos}, {Drazinos}, {Dur{\'a}n}, {Dzigan}, {Ecale}, {Edvardsson}, {Enke},
  {Erdmann}, {Escolar}, {Espina}, {Evans}, {Eynard Bontemps}, {Fabre},
  {Fabrizio}, {Faigler}, {Falc{\~a}o}, {Farr{\`a}s Casas}, {Faye}, {Federici},
  {Fedorets}, {Fern{\'a}ndez-Hern{\'a}ndez}, {Fernique}, {Fienga}, {Figueras},
  {Filippi}, {Findeisen}, {Fonti}, {Fouesneau}, {Fraile}, {Fraser}, {Fuchs},
  {Furnell}, {Gai}, {Galleti}, {Galluccio}, {Garabato}, {Garc{\'\i}a-Sedano},
  {Gar{\'e}}, {Garofalo}, {Garralda}, {Gavras}, {Gerssen}, {Geyer}, {Gilmore},
  {Girona}, {Giuffrida}, {Gomes}, {Gonz{\'a}lez-Marcos},
  {Gonz{\'a}lez-N{\'u}{\~n}ez}, {Gonz{\'a}lez-Vidal}, {Granvik}, {Guerrier},
  {Guillout}, {Guiraud}, {G{\'u}rpide}, {Guti{\'e}rrez-S{\'a}nchez}, {Guy},
  {Haigron}, {Hatzidimitriou}, {Haywood}, {Heiter}, {Helmi}, {Hobbs},
  {Hofmann}, {Holl}, {Holland}, {Hunt}, {Hypki}, {Icardi}, {Irwin}, {Jevardat
  de Fombelle}, {Jofr{\'e}}, {Jonker}, {Jorissen}, {Julbe}, {Karampelas},
  {Kochoska}, {Kohley}, {Kolenberg}, {Kontizas}, {Koposov}, {Kordopatis},
  {Koubsky}, {Kowalczyk}, {Krone-Martins}, {Kudryashova}, {Kull}, {Bachchan},
  {Lacoste-Seris}, {Lanza}, {Lavigne}, {Le Poncin-Lafitte}, {Lebreton},
  {Lebzelter}, {Leccia}, {Leclerc}, {Lecoeur-Taibi}, {Lemaitre}, {Lenhardt},
  {Leroux}, {Liao}, {Licata}, {Lindstr{\o}m}, {Lister}, {Livanou}, {Lobel},
  {L{\"o}ffler}, {L{\'o}pez}, {Lopez-Lozano}, {Lorenz}, {Loureiro},
  {MacDonald}, {Magalh{\~a}es Fernandes}, {Managau}, {Mann}, {Mantelet},
  {Marchal}, {Marchant}, {Marconi}, {Marie}, {Marinoni}, {Marrese},
  {Marschalk{\'o}}, {Marshall}, {Mart{\'\i}n-Fleitas}, {Martino}, {Mary},
  {Matijevi{\v{c}}}, {Mazeh}, {McMillan}, {Messina}, {Mestre}, {Michalik},
  {Millar}, {Miranda}, {Molina}, {Molinaro}, {Molinaro}, {Moln{\'a}r},
  {Moniez}, {Montegriffo}, {Monteiro}, {Mor}, {Mora}, {Morbidelli}, {Morel},
  {Morgenthaler}, {Morley}, {Morris}, {Mulone}, {Muraveva}, {Musella},
  {Narbonne}, {Nelemans}, {Nicastro}, {Noval}, {Ord{\'e}novic},
  {Ordieres-Mer{\'e}}, {Osborne}, {Pagani}, {Pagano}, {Pailler}, {Palacin},
  {Palaversa}, {Parsons}, {Paulsen}, {Pecoraro}, {Pedrosa}, {Pentik{\"a}inen},
  {Pereira}, {Pichon}, {Piersimoni}, {Pineau}, {Plachy}, {Plum}, {Poujoulet},
  {Pr{\v{s}}a}, {Pulone}, {Ragaini}, {Rago}, {Rambaux}, {Ramos-Lerate},
  {Ranalli}, {Rauw}, {Read}, {Regibo}, {Renk}, {Reyl{\'e}}, {Ribeiro},
  {Rimoldini}, {Ripepi}, {Riva}, {Rixon}, {Roelens}, {Romero-G{\'o}mez},
  {Rowell}, {Royer}, {Rudolph}, {Ruiz-Dern}, {Sadowski}, {Sagrist{\`a}
  Sell{\'e}s}, {Sahlmann}, {Salgado}, {Salguero}, {Sarasso}, {Savietto},
  {Schnorhk}, {Schultheis}, {Sciacca}, {Segol}, {Segovia}, {Segransan},
  {Serpell}, {Shih}, {Smareglia}, {Smart}, {Smith}, {Solano}, {Solitro},
  {Sordo}, {Soria Nieto}, {Souchay}, {Spagna}, {Spoto}, {Stampa}, {Steele},
  {Steidelm{\"u}ller}, {Stephenson}, {Stoev}, {Suess}, {S{\"u}veges}, {Surdej},
  {Szabados}, {Szegedi-Elek}, {Tapiador}, {Taris}, {Tauran}, {Taylor},
  {Teixeira}, {Terrett}, {Tingley}, {Trager}, {Turon}, {Ulla}, {Utrilla},
  {Valentini}, {van Elteren}, {Van Hemelryck}, {van Leeuwen}, {Varadi},
  {Vecchiato}, {Veljanoski}, {Via}, {Vicente}, {Vogt}, {Voss}, {Votruba},
  {Voutsinas}, {Walmsley}, {Weiler}, {Weingrill}, {Werner}, {Wevers},
  {Whitehead}, {Wyrzykowski}, {Yoldas}, {{\v{Z}}erjal}, {Zucker}, {Zurbach},
  {Zwitter}, {Alecu}, {Allen}, {Allende Prieto}, {Amorim},
  {Anglada-Escud{\'e}}, {Arsenijevic}, {Azaz}, {Balm}, {Beck}, {Bernstein},
  {Bigot}, {Bijaoui}, {Blasco}, {Bonfigli}, {Bono}, {Boudreault}, {Bressan},
  {Brown}, {Brunet}, {Bunclark}, {Buonanno}, {Butkevich}, {Carret}, {Carrion},
  {Chemin}, {Ch{\'e}reau}, {Corcione}, {Darmigny}, {de Boer}, {de Teodoro}, {de
  Zeeuw}, {Delle Luche}, {Domingues}, {Dubath}, {Fodor}, {Fr{\'e}zouls},
  {Fries}, {Fustes}, {Fyfe}, {Gallardo}, {Gallegos}, {Gardiol}, {Gebran},
  {Gomboc}, {G{\'o}mez}, {Grux}, {Gueguen}, {Heyrovsky}, {Hoar}, {Iannicola},
  {Isasi Parache}, {Janotto}, {Joliet}, {Jonckheere}, {Keil}, {Kim},
  {Klagyivik}, {Klar}, {Knude}, {Kochukhov}, {Kolka}, {Kos}, {Kutka}, {Lainey},
  {LeBouquin}, {Liu}, {Loreggia}, {Makarov}, {Marseille}, {Martayan},
  {Martinez-Rubi}, {Massart}, {Meynadier}, {Mignot}, {Munari}, {Nguyen},
  {Nordlander}, {Ocvirk}, {O'Flaherty}, {Olias Sanz}, {Ortiz}, {Osorio},
  {Oszkiewicz}, {Ouzounis}, {Palmer}, {Park}, {Pasquato}, {Peltzer}, {Peralta},
  {P{\'e}turaud}, {Pieniluoma}, {Pigozzi}, {Poels}, {Prat}, {Prod'homme},
  {Raison}, {Rebordao}, {Risquez}, {Rocca-Volmerange}, {Rosen}, {Ruiz-Fuertes},
  {Russo}, {Sembay}, {Serraller Vizcaino}, {Short}, {Siebert}, {Silva},
  {Sinachopoulos}, {Slezak}, {Soffel}, {Sosnowska}, {Strai{\v{z}}ys}, {ter
  Linden}, {Terrell}, {Theil}, {Tiede}, {Troisi}, {Tsalmantza}, {Tur},
  {Vaccari}, {Vachier}, {Valles}, {Van Hamme}, {Veltz}, {Virtanen}, {Wallut},
  {Wichmann}, {Wilkinson}, {Ziaeepour}, \& {Zschocke}}]{2016A&A...595A...1G}
{Gaia Collaboration}, {Prusti}, T., {de Bruijne}, J.~H.~J., {et~al.} 2016,
  \aap, 595, A1, \dodoi{10.1051/0004-6361/201629272}

\bibitem[{{Gaia Collaboration} {et~al.}(2023){Gaia Collaboration}, {Vallenari},
  {Brown}, {Prusti}, {de Bruijne}, {Arenou}, {Babusiaux}, {Biermann},
  {Creevey}, {Ducourant}, {Evans}, {Eyer}, {Guerra}, {Hutton}, {Jordi},
  {Klioner}, {Lammers}, {Lindegren}, {Luri}, {Mignard}, {Panem}, {Pourbaix},
  {Randich}, {Sartoretti}, {Soubiran}, {Tanga}, {Walton}, {Bailer-Jones},
  {Bastian}, {Drimmel}, {Jansen}, {Katz}, {Lattanzi}, {van Leeuwen}, {Bakker},
  {Cacciari}, {Casta{\~n}eda}, {De Angeli}, {Fabricius}, {Fouesneau},
  {Fr{\'e}mat}, {Galluccio}, {Guerrier}, {Heiter}, {Masana}, {Messineo},
  {Mowlavi}, {Nicolas}, {Nienartowicz}, {Pailler}, {Panuzzo}, {Riclet}, {Roux},
  {Seabroke}, {Sordo}, {Th{\'e}venin}, {Gracia-Abril}, {Portell}, {Teyssier},
  {Altmann}, {Andrae}, {Audard}, {Bellas-Velidis}, {Benson}, {Berthier},
  {Blomme}, {Burgess}, {Busonero}, {Busso}, {C{\'a}novas}, {Carry}, {Cellino},
  {Cheek}, {Clementini}, {Damerdji}, {Davidson}, {de Teodoro}, {Nu{\~n}ez
  Campos}, {Delchambre}, {Dell'Oro}, {Esquej}, {Fern{\'a}ndez-Hern{\'a}ndez},
  {Fraile}, {Garabato}, {Garc{\'\i}a-Lario}, {Gosset}, {Haigron}, {Halbwachs},
  {Hambly}, {Harrison}, {Hern{\'a}ndez}, {Hestroffer}, {Hodgkin}, {Holl},
  {Jan{\ss}en}, {Jevardat de Fombelle}, {Jordan}, {Krone-Martins}, {Lanzafame},
  {L{\"o}ffler}, {Marchal}, {Marrese}, {Moitinho}, {Muinonen}, {Osborne},
  {Pancino}, {Pauwels}, {Recio-Blanco}, {Reyl{\'e}}, {Riello}, {Rimoldini},
  {Roegiers}, {Rybizki}, {Sarro}, {Siopis}, {Smith}, {Sozzetti}, {Utrilla},
  {van Leeuwen}, {Abbas}, {{\'A}brah{\'a}m}, {Abreu Aramburu}, {Aerts},
  {Aguado}, {Ajaj}, {Aldea-Montero}, {Altavilla}, {{\'A}lvarez}, {Alves},
  {Anders}, {Anderson}, {Anglada Varela}, {Antoja}, {Baines}, {Baker},
  {Balaguer-N{\'u}{\~n}ez}, {Balbinot}, {Balog}, {Barache}, {Barbato},
  {Barros}, {Barstow}, {Bartolom{\'e}}, {Bassilana}, {Bauchet}, {Becciani},
  {Bellazzini}, {Berihuete}, {Bernet}, {Bertone}, {Bianchi}, {Binnenfeld},
  {Blanco-Cuaresma}, {Blazere}, {Boch}, {Bombrun}, {Bossini}, {Bouquillon},
  {Bragaglia}, {Bramante}, {Breedt}, {Bressan}, {Brouillet}, {Brugaletta},
  {Bucciarelli}, {Burlacu}, {Butkevich}, {Buzzi}, {Caffau}, {Cancelliere},
  {Cantat-Gaudin}, {Carballo}, {Carlucci}, {Carnerero}, {Carrasco},
  {Casamiquela}, {Castellani}, {Castro-Ginard}, {Chaoul}, {Charlot}, {Chemin},
  {Chiaramida}, {Chiavassa}, {Chornay}, {Comoretto}, {Contursi}, {Cooper},
  {Cornez}, {Cowell}, {Crifo}, {Cropper}, {Crosta}, {Crowley}, {Dafonte},
  {Dapergolas}, {David}, {David}, {de Laverny}, {De Luise}, {De March}, {De
  Ridder}, {de Souza}, {de Torres}, {del Peloso}, {del Pozo}, {Delbo},
  {Delgado}, {Delisle}, {Demouchy}, {Dharmawardena}, {Di Matteo}, {Diakite},
  {Diener}, {Distefano}, {Dolding}, {Edvardsson}, {Enke}, {Fabre}, {Fabrizio},
  {Faigler}, {Fedorets}, {Fernique}, {Fienga}, {Figueras}, {Fournier},
  {Fouron}, {Fragkoudi}, {Gai}, {Garcia-Gutierrez}, {Garcia-Reinaldos},
  {Garc{\'\i}a-Torres}, {Garofalo}, {Gavel}, {Gavras}, {Gerlach}, {Geyer},
  {Giacobbe}, {Gilmore}, {Girona}, {Giuffrida}, {Gomel}, {Gomez},
  {Gonz{\'a}lez-N{\'u}{\~n}ez}, {Gonz{\'a}lez-Santamar{\'\i}a},
  {Gonz{\'a}lez-Vidal}, {Granvik}, {Guillout}, {Guiraud},
  {Guti{\'e}rrez-S{\'a}nchez}, {Guy}, {Hatzidimitriou}, {Hauser}, {Haywood},
  {Helmer}, {Helmi}, {Sarmiento}, {Hidalgo}, {Hilger}, {H{\l}adczuk}, {Hobbs},
  {Holland}, {Huckle}, {Jardine}, {Jasniewicz}, {Jean-Antoine Piccolo},
  {Jim{\'e}nez-Arranz}, {Jorissen}, {Juaristi Campillo}, {Julbe}, {Karbevska},
  {Kervella}, {Khanna}, {Kontizas}, {Kordopatis}, {Korn}, {K{\'o}sp{\'a}l},
  {Kostrzewa-Rutkowska}, {Kruszy{\'n}ska}, {Kun}, {Laizeau}, {Lambert},
  {Lanza}, {Lasne}, {Le Campion}, {Lebreton}, {Lebzelter}, {Leccia}, {Leclerc},
  {Lecoeur-Taibi}, {Liao}, {Licata}, {Lindstr{\o}m}, {Lister}, {Livanou},
  {Lobel}, {Lorca}, {Loup}, {Madrero Pardo}, {Magdaleno Romeo}, {Managau},
  {Mann}, {Manteiga}, {Marchant}, {Marconi}, {Marcos}, {Marcos Santos},
  {Mar{\'\i}n Pina}, {Marinoni}, {Marocco}, {Marshall}, {Martin Polo},
  {Mart{\'\i}n-Fleitas}, {Marton}, {Mary}, {Masip}, {Massari},
  {Mastrobuono-Battisti}, {Mazeh}, {McMillan}, {Messina}, {Michalik}, {Millar},
  {Mints}, {Molina}, {Molinaro}, {Moln{\'a}r}, {Monari}, {Mongui{\'o}},
  {Montegriffo}, {Montero}, {Mor}, {Mora}, {Morbidelli}, {Morel}, {Morris},
  {Muraveva}, {Murphy}, {Musella}, {Nagy}, {Noval}, {Oca{\~n}a}, {Ogden},
  {Ordenovic}, {Osinde}, {Pagani}, {Pagano}, {Palaversa}, {Palicio},
  {Pallas-Quintela}, {Panahi}, {Payne-Wardenaar}, {Pe{\~n}alosa Esteller},
  {Penttil{\"a}}, {Pichon}, {Piersimoni}, {Pineau}, {Plachy}, {Plum}, {Poggio},
  {Pr{\v{s}}a}, {Pulone}, {Racero}, {Ragaini}, {Rainer}, {Raiteri}, {Rambaux},
  {Ramos}, {Ramos-Lerate}, {Re Fiorentin}, {Regibo}, {Richards}, {Rios Diaz},
  {Ripepi}, {Riva}, {Rix}, {Rixon}, {Robichon}, {Robin}, {Robin}, {Roelens},
  {Rogues}, {Rohrbasser}, {Romero-G{\'o}mez}, {Rowell}, {Royer}, {Ruz Mieres},
  {Rybicki}, {Sadowski}, {S{\'a}ez N{\'u}{\~n}ez}, {Sagrist{\`a} Sell{\'e}s},
  {Sahlmann}, {Salguero}, {Samaras}, {Sanchez Gimenez}, {Sanna},
  {Santove{\~n}a}, {Sarasso}, {Schultheis}, {Sciacca}, {Segol}, {Segovia},
  {S{\'e}gransan}, {Semeux}, {Shahaf}, {Siddiqui}, {Siebert}, {Siltala},
  {Silvelo}, {Slezak}, {Slezak}, {Smart}, {Snaith}, {Solano}, {Solitro},
  {Souami}, {Souchay}, {Spagna}, {Spina}, {Spoto}, {Steele},
  {Steidelm{\"u}ller}, {Stephenson}, {S{\"u}veges}, {Surdej}, {Szabados},
  {Szegedi-Elek}, {Taris}, {Taylor}, {Teixeira}, {Tolomei}, {Tonello}, {Torra},
  {Torra}, {Torralba Elipe}, {Trabucchi}, {Tsounis}, {Turon}, {Ulla}, {Unger},
  {Vaillant}, {van Dillen}, {van Reeven}, {Vanel}, {Vecchiato}, {Viala},
  {Vicente}, {Voutsinas}, {Weiler}, {Wevers}, {Wyrzykowski}, {Yoldas}, {Yvard},
  {Zhao}, {Zorec}, {Zucker}, \& {Zwitter}}]{2023A&A...674A...1G}
{Gaia Collaboration}, {Vallenari}, A., {Brown}, A.~G.~A., {et~al.} 2023, \aap,
  674, A1, \dodoi{10.1051/0004-6361/202243940}

\bibitem[{{Gavazzi} {et~al.}(2015){Gavazzi}, {Consolandi}, {Dotti}, {Fanali},
  {Fossati}, {Fumagalli}, {Viscardi}, {Savorgnan}, {Boselli}, {Guti{\'e}rrez},
  {Hern{\'a}ndez Toledo}, {Giovanelli}, \& {Haynes}}]{2015A&A...580A.116G}
{Gavazzi}, G., {Consolandi}, G., {Dotti}, M., {et~al.} 2015, \aap, 580, A116,
  \dodoi{10.1051/0004-6361/201425351}

\bibitem[{{G{\'e}ron} {et~al.}(2021){G{\'e}ron}, {Smethurst}, {Lintott},
  {Kruk}, {Masters}, {Simmons}, \& {Stark}}]{2021MNRAS.507.4389G}
{G{\'e}ron}, T., {Smethurst}, R.~J., {Lintott}, C., {et~al.} 2021, \mnras, 507,
  4389, \dodoi{10.1093/mnras/stab2064}

\bibitem[{{Grimes} {et~al.}(2006){Grimes}, {Heckman}, {Hoopes}, {Strickland},
  {Aloisi}, {Meurer}, \& {Ptak}}]{2006ApJ...648..310G}
{Grimes}, J.~P., {Heckman}, T., {Hoopes}, C., {et~al.} 2006, \apj, 648, 310,
  \dodoi{10.1086/505680}

\bibitem[{{Guo} {et~al.}(2023){Guo}, {Jogee}, {Finkelstein}, {Chen}, {Wise},
  {Bagley}, {Barro}, {Wuyts}, {Kocevski}, {Kartaltepe}, {McGrath}, {Ferguson},
  {Mobasher}, {Giavalisco}, {Lucas}, {Zavala}, {Lotz}, {Grogin},
  {Huertas-Company}, {Vega-Ferrero}, {Hathi}, {Haro}, {Dickinson}, {Koekemoer},
  {Papovich}, {Pirzkal}, {Yung}, {Backhaus}, {Bell}, {Calabr{\`o}}, {Cleri},
  {Coogan}, {Cooper}, {Costantin}, {Croton}, {Davis}, {Dekel}, {Franco},
  {Gardner}, {Holwerda}, {Hutchison}, {Pandya}, {P{\'e}rez-Gonz{\'a}lez},
  {Ravindranath}, {Rose}, {Trump}, {de la Vega}, \&
  {Wang}}]{2023ApJ...945L..10G}
{Guo}, Y., {Jogee}, S., {Finkelstein}, S.~L., {et~al.} 2023, \apjl, 945, L10,
  \dodoi{10.3847/2041-8213/acacfb}

\bibitem[{{Harris} {et~al.}(2012){Harris}, {Baker}, {Frayer}, {Smail},
  {Swinbank}, {Riechers}, {van der Werf}, {Auld}, {Baes}, {Bussmann},
  {Buttiglione}, {Cava}, {Clements}, {Cooray}, {Dannerbauer}, {Dariush}, {De
  Zotti}, {Dunne}, {Dye}, {Eales}, {Fritz}, {Gonz{\'a}lez-Nuevo}, {Hopwood},
  {Ibar}, {Ivison}, {Jarvis}, {Maddox}, {Negrello}, {Rigby}, {Smith}, {Temi},
  \& {Wardlow}}]{2012ApJ...752..152H}
{Harris}, A.~I., {Baker}, A.~J., {Frayer}, D.~T., {et~al.} 2012, \apj, 752,
  152, \dodoi{10.1088/0004-637X/752/2/152}

\bibitem[{Harris {et~al.}(2020)Harris, Millman, van~der Walt, Gommers,
  Virtanen, Cournapeau, Wieser, Taylor, Berg, Smith, Kern, Picus, Hoyer, van
  Kerkwijk, Brett, Haldane, del R{\'{i}}o, Wiebe, Peterson,
  G{\'{e}}rard-Marchant, Sheppard, Reddy, Weckesser, Abbasi, Gohlke, \&
  Oliphant}]{harris2020array}
Harris, C.~R., Millman, K.~J., van~der Walt, S.~J., {et~al.} 2020, Nature, 585,
  357, \dodoi{10.1038/s41586-020-2649-2}

\bibitem[{{Hayward} {et~al.}(2012){Hayward}, {Jonsson}, {Kere{\v{s}}},
  {Magnelli}, {Hernquist}, \& {Cox}}]{2012MNRAS.424..951H}
{Hayward}, C.~C., {Jonsson}, P., {Kere{\v{s}}}, D., {et~al.} 2012, \mnras, 424,
  951, \dodoi{10.1111/j.1365-2966.2012.21254.x}

\bibitem[{{Hayward} {et~al.}(2011){Hayward}, {Kere{\v{s}}}, {Jonsson},
  {Narayanan}, {Cox}, \& {Hernquist}}]{2011ApJ...743..159H}
{Hayward}, C.~C., {Kere{\v{s}}}, D., {Jonsson}, P., {et~al.} 2011, \apj, 743,
  159, \dodoi{10.1088/0004-637X/743/2/159}

\bibitem[{{Hayward} {et~al.}(2013){Hayward}, {Narayanan}, {Kere{\v{s}}},
  {Jonsson}, {Hopkins}, {Cox}, \& {Hernquist}}]{2013MNRAS.428.2529H}
{Hayward}, C.~C., {Narayanan}, D., {Kere{\v{s}}}, D., {et~al.} 2013, \mnras,
  428, 2529, \dodoi{10.1093/mnras/sts222}

\bibitem[{{Ho}(2007)}]{2007ApJ...668...94H}
{Ho}, L.~C. 2007, \apj, 668, 94, \dodoi{10.1086/521270}

\bibitem[{{Hodge} {et~al.}(2019){Hodge}, {Smail}, {Walter}, {da Cunha},
  {Swinbank}, {Rybak}, {Venemans}, {Brandt}, {Calistro Rivera}, {Chapman},
  {Chen}, {Cox}, {Dannerbauer}, {Decarli}, {Greve}, {Knudsen}, {Menten},
  {Schinnerer}, {Simpson}, {van der Werf}, {Wardlow}, \&
  {Weiss}}]{2019ApJ...876..130H}
{Hodge}, J.~A., {Smail}, I., {Walter}, F., {et~al.} 2019, \apj, 876, 130,
  \dodoi{10.3847/1538-4357/ab1846}

\bibitem[{{Hughes} {et~al.}(1998){Hughes}, {Serjeant}, {Dunlop},
  {Rowan-Robinson}, {Blain}, {Mann}, {Ivison}, {Peacock}, {Efstathiou}, {Gear},
  {Oliver}, {Lawrence}, {Longair}, {Goldschmidt}, \&
  {Jenness}}]{1998Natur.394..241H}
{Hughes}, D.~H., {Serjeant}, S., {Dunlop}, J., {et~al.} 1998, \nat, 394, 241,
  \dodoi{10.1038/28328}

\bibitem[{Hunter(2007)}]{Hunter:2007}
Hunter, J.~D. 2007, Computing in Science \& Engineering, 9, 90,
  \dodoi{10.1109/MCSE.2007.55}

\bibitem[{{Kennicutt}(1981)}]{1981AJ.....86.1847K}
{Kennicutt}, R.~C., J. 1981, \aj, 86, 1847, \dodoi{10.1086/113064}

\bibitem[{{Kere{\v{s}}} {et~al.}(2009){Kere{\v{s}}}, {Katz}, {Fardal},
  {Dav{\'e}}, \& {Weinberg}}]{2009MNRAS.395..160K}
{Kere{\v{s}}}, D., {Katz}, N., {Fardal}, M., {Dav{\'e}}, R., \& {Weinberg},
  D.~H. 2009, \mnras, 395, 160, \dodoi{10.1111/j.1365-2966.2009.14541.x}

\bibitem[{{Kim} {et~al.}(2017){Kim}, {Hwang}, {Chung}, {Lee}, {Park},
  {Cervantes Sodi}, \& {Kim}}]{2017ApJ...845...93K}
{Kim}, E., {Hwang}, H.~S., {Chung}, H., {et~al.} 2017, \apj, 845, 93,
  \dodoi{10.3847/1538-4357/aa80db}

\bibitem[{{Kuno} {et~al.}(2007){Kuno}, {Sato}, {Nakanishi}, {Hirota}, {Tosaki},
  {Shioya}, {Sorai}, {Nakai}, {Nishiyama}, \&
  {Vila-Vilar{\'o}}}]{2007PASJ...59..117K}
{Kuno}, N., {Sato}, N., {Nakanishi}, H., {et~al.} 2007, \pasj, 59, 117,
  \dodoi{10.1093/pasj/59.1.117}

\bibitem[{{Leja} {et~al.}(2019){Leja}, {Johnson}, {Conroy}, {van Dokkum},
  {Speagle}, {Brammer}, {Momcheva}, {Skelton}, {Whitaker}, {Franx}, \&
  {Nelson}}]{2019ApJ...877..140L}
{Leja}, J., {Johnson}, B.~D., {Conroy}, C., {et~al.} 2019, \apj, 877, 140,
  \dodoi{10.3847/1538-4357/ab1d5a}

\bibitem[{{Lin} {et~al.}(2020){Lin}, {Li}, {Du}, {Wang}, {Xiao}, {Bureau},
  {Fraser-McKelvie}, {Masters}, {Lin}, {Wake}, \& {Hao}}]{2020MNRAS.499.1406L}
{Lin}, L., {Li}, C., {Du}, C., {et~al.} 2020, \mnras, 499, 1406,
  \dodoi{10.1093/mnras/staa2913}

\bibitem[{{Linden} {et~al.}(2023){Linden}, {Evans}, {Armus}, {Rich}, {Larson},
  {Lai}, {Privon}, {U}, {Inami}, {Bohn}, {Song}, {Barcos-Mu{\~n}oz},
  {Charmandaris}, {Medling}, {Stierwalt}, {Diaz-Santos}, {B{\"o}ker}, {van der
  Werf}, {Aalto}, {Appleton}, {Brown}, {Hayward}, {Howell}, {Iwasawa},
  {Kemper}, {Frayer}, {Law}, {Malkan}, {Marshall}, {Mazzarella}, {Murphy},
  {Sanders}, \& {Surace}}]{2023ApJ...944L..55L}
{Linden}, S.~T., {Evans}, A.~S., {Armus}, L., {et~al.} 2023, \apjl, 944, L55,
  \dodoi{10.3847/2041-8213/acb335}

\bibitem[{{Lovell} {et~al.}(2021){Lovell}, {Geach}, {Dav{\'e}}, {Narayanan}, \&
  {Li}}]{2021MNRAS.502..772L}
{Lovell}, C.~C., {Geach}, J.~E., {Dav{\'e}}, R., {Narayanan}, D., \& {Li}, Q.
  2021, \mnras, 502, 772, \dodoi{10.1093/mnras/staa4043}

\bibitem[{{Madau} \& {Dickinson}(2014)}]{2014ARA&A..52..415M}
{Madau}, P., \& {Dickinson}, M. 2014, \araa, 52, 415,
  \dodoi{10.1146/annurev-astro-081811-125615}

\bibitem[{{Maeda} {et~al.}(2023){Maeda}, {Egusa}, {Ohta}, {Fujimoto}, \&
  {Habe}}]{2023ApJ...943....7M}
{Maeda}, F., {Egusa}, F., {Ohta}, K., {Fujimoto}, Y., \& {Habe}, A. 2023, \apj,
  943, 7, \dodoi{10.3847/1538-4357/aca664}

\bibitem[{{Miettinen} {et~al.}(2017){Miettinen}, {Delvecchio},
  {Smol{\v{c}}i{\'c}}, {Aravena}, {Brisbin}, {Karim}, {Magnelli}, {Novak},
  {Schinnerer}, {Albrecht}, {Aussel}, {Bertoldi}, {Capak}, {Casey}, {Hayward},
  {Ilbert}, {Intema}, {Jiang}, {Le F{\`e}vre}, {McCracken}, {Mu{\~n}oz
  Arancibia}, {Navarrete}, {Padilla}, {Riechers}, {Salvato}, {Scott}, {Sheth},
  \& {Tasca}}]{2017A&A...606A..17M}
{Miettinen}, O., {Delvecchio}, I., {Smol{\v{c}}i{\'c}}, V., {et~al.} 2017,
  \aap, 606, A17, \dodoi{10.1051/0004-6361/201730762}

\bibitem[{{Mizukoshi} {et~al.}(2021){Mizukoshi}, {Kohno}, {Egusa}, {Hatsukade},
  {Minezaki}, {Saito}, {Tamura}, {Iono}, {Ueda}, {Matsuda}, {Kawabe}, {Lee},
  {Yun}, \& {Espada}}]{2021ApJ...917...94M}
{Mizukoshi}, S., {Kohno}, K., {Egusa}, F., {et~al.} 2021, \apj, 917, 94,
  \dodoi{10.3847/1538-4357/ac01cc}

\bibitem[{{Narayanan} {et~al.}(2010){Narayanan}, {Hayward}, {Cox}, {Hernquist},
  {Jonsson}, {Younger}, \& {Groves}}]{2010MNRAS.401.1613N}
{Narayanan}, D., {Hayward}, C.~C., {Cox}, T.~J., {et~al.} 2010, \mnras, 401,
  1613, \dodoi{10.1111/j.1365-2966.2009.15790.x}

\bibitem[{{Narayanan} {et~al.}(2015){Narayanan}, {Turk}, {Feldmann},
  {Robitaille}, {Hopkins}, {Thompson}, {Hayward}, {Ball},
  {Faucher-Gigu{\`e}re}, \& {Kere{\v{s}}}}]{2015Natur.525..496N}
{Narayanan}, D., {Turk}, M., {Feldmann}, R., {et~al.} 2015, \nat, 525, 496,
  \dodoi{10.1038/nature15383}

\bibitem[{{Noll} {et~al.}(2009){Noll}, {Burgarella}, {Giovannoli}, {Buat},
  {Marcillac}, \& {Mu{\~n}oz-Mateos}}]{2009A&A...507.1793N}
{Noll}, S., {Burgarella}, D., {Giovannoli}, E., {et~al.} 2009, \aap, 507, 1793,
  \dodoi{10.1051/0004-6361/200912497}

\bibitem[{{Oh} {et~al.}(2008){Oh}, {Kim}, {Lee}, \&
  {Kim}}]{2008ApJ...683...94O}
{Oh}, S.~H., {Kim}, W.-T., {Lee}, H.~M., \& {Kim}, J. 2008, \apj, 683, 94,
  \dodoi{10.1086/588184}

\bibitem[{{Planck Collaboration} {et~al.}(2020){Planck Collaboration},
  {Aghanim}, {Akrami}, {Arroja}, {Ashdown}, {Aumont}, {Baccigalupi},
  {Ballardini}, {Banday}, {Barreiro}, {Bartolo}, {Basak}, {Battye}, {Benabed},
  {Bernard}, {Bersanelli}, {Bielewicz}, {Bock}, {Bond}, {Borrill}, {Bouchet},
  {Boulanger}, {Bucher}, {Burigana}, {Butler}, {Calabrese}, {Cardoso},
  {Carron}, {Casaponsa}, {Challinor}, {Chiang}, {Colombo}, {Combet},
  {Contreras}, {Crill}, {Cuttaia}, {de Bernardis}, {de Zotti}, {Delabrouille},
  {Delouis}, {D{\'e}sert}, {Di Valentino}, {Dickinson}, {Diego}, {Donzelli},
  {Dor{\'e}}, {Douspis}, {Ducout}, {Dupac}, {Efstathiou}, {Elsner},
  {En{\ss}lin}, {Eriksen}, {Falgarone}, {Fantaye}, {Fergusson},
  {Fernandez-Cobos}, {Finelli}, {Forastieri}, {Frailis}, {Franceschi},
  {Frolov}, {Galeotta}, {Galli}, {Ganga}, {G{\'e}nova-Santos}, {Gerbino},
  {Ghosh}, {Gonz{\'a}lez-Nuevo}, {G{\'o}rski}, {Gratton}, {Gruppuso},
  {Gudmundsson}, {Hamann}, {Handley}, {Hansen}, {Helou}, {Herranz},
  {Hildebrandt}, {Hivon}, {Huang}, {Jaffe}, {Jones}, {Karakci}, {Keih{\"a}nen},
  {Keskitalo}, {Kiiveri}, {Kim}, {Kisner}, {Knox}, {Krachmalnicoff}, {Kunz},
  {Kurki-Suonio}, {Lagache}, {Lamarre}, {Langer}, {Lasenby}, {Lattanzi},
  {Lawrence}, {Le Jeune}, {Leahy}, {Lesgourgues}, {Levrier}, {Lewis},
  {Liguori}, {Lilje}, {Lilley}, {Lindholm}, {L{\'o}pez-Caniego}, {Lubin}, {Ma},
  {Mac{\'\i}as-P{\'e}rez}, {Maggio}, {Maino}, {Mandolesi}, {Mangilli},
  {Marcos-Caballero}, {Maris}, {Martin}, {Martinelli},
  {Mart{\'\i}nez-Gonz{\'a}lez}, {Matarrese}, {Mauri}, {McEwen}, {Meerburg},
  {Meinhold}, {Melchiorri}, {Mennella}, {Migliaccio}, {Millea}, {Mitra},
  {Miville-Desch{\^e}nes}, {Molinari}, {Moneti}, {Montier}, {Morgante}, {Moss},
  {Mottet}, {M{\"u}nchmeyer}, {Natoli}, {N{\o}rgaard-Nielsen}, {Oxborrow},
  {Pagano}, {Paoletti}, {Partridge}, {Patanchon}, {Pearson}, {Peel}, {Peiris},
  {Perrotta}, {Pettorino}, {Piacentini}, {Polastri}, {Polenta}, {Puget},
  {Rachen}, {Reinecke}, {Remazeilles}, {Renault}, {Renzi}, {Rocha}, {Rosset},
  {Roudier}, {Rubi{\~n}o-Mart{\'\i}n}, {Ruiz-Granados}, {Salvati}, {Sandri},
  {Savelainen}, {Scott}, {Shellard}, {Shiraishi}, {Sirignano}, {Sirri},
  {Spencer}, {Sunyaev}, {Suur-Uski}, {Tauber}, {Tavagnacco}, {Tenti},
  {Terenzi}, {Toffolatti}, {Tomasi}, {Trombetti}, {Valiviita}, {Van Tent},
  {Vibert}, {Vielva}, {Villa}, {Vittorio}, {Wandelt}, {Wehus}, {White},
  {White}, {Zacchei}, \& {Zonca}}]{2020A&A...641A...1P}
{Planck Collaboration}, {Aghanim}, N., {Akrami}, Y., {et~al.} 2020, \aap, 641,
  A1, \dodoi{10.1051/0004-6361/201833880}

\bibitem[{{Rosas-Guevara} {et~al.}(2022){Rosas-Guevara}, {Bonoli}, {Dotti},
  {Izquierdo-Villalba}, {Lupi}, {Zana}, {Bonetti}, {Nelson}, {Springel},
  {Hernquist}, \& {Vogelsberger}}]{2022MNRAS.512.5339R}
{Rosas-Guevara}, Y., {Bonoli}, S., {Dotti}, M., {et~al.} 2022, \mnras, 512,
  5339, \dodoi{10.1093/mnras/stac816}

\bibitem[{{Sanders} \& {Mirabel}(1996)}]{1996ARA&A..34..749S}
{Sanders}, D.~B., \& {Mirabel}, I.~F. 1996, \araa, 34, 749,
  \dodoi{10.1146/annurev.astro.34.1.749}

\bibitem[{{Simmons} {et~al.}(2014){Simmons}, {Melvin}, {Lintott}, {Masters},
  {Willett}, {Keel}, {Smethurst}, {Cheung}, {Nichol}, {Schawinski},
  {Rutkowski}, {Kartaltepe}, {Bell}, {Casteels}, {Conselice}, {Almaini},
  {Ferguson}, {Fortson}, {Hartley}, {Kocevski}, {Koekemoer}, {McIntosh},
  {Mortlock}, {Newman}, {Ownsworth}, {Bamford}, {Dahlen}, {Faber},
  {Finkelstein}, {Fontana}, {Galametz}, {Grogin}, {Gr{\"u}tzbauch}, {Guo},
  {H{\"a}u{\ss}ler}, {Jek}, {Kaviraj}, {Lucas}, {Peth}, {Salvato}, {Wiklind},
  \& {Wuyts}}]{2014MNRAS.445.3466S}
{Simmons}, B.~D., {Melvin}, T., {Lintott}, C., {et~al.} 2014, \mnras, 445,
  3466, \dodoi{10.1093/mnras/stu1817}

\bibitem[{{Simons} {et~al.}(2017){Simons}, {Kassin}, {Weiner}, {Faber},
  {Trump}, {Heckman}, {Koo}, {Pacifici}, {Primack}, {Snyder}, \& {de la
  Vega}}]{2017ApJ...843...46S}
{Simons}, R.~C., {Kassin}, S.~A., {Weiner}, B.~J., {et~al.} 2017, \apj, 843,
  46, \dodoi{10.3847/1538-4357/aa740c}

\bibitem[{{Smail} {et~al.}(1997){Smail}, {Ivison}, \&
  {Blain}}]{1997ApJ...490L...5S}
{Smail}, I., {Ivison}, R.~J., \& {Blain}, A.~W. 1997, \apjl, 490, L5,
  \dodoi{10.1086/311017}

\bibitem[{{Smail} {et~al.}(2023){Smail}, {Dudzeviciute}, {Gurwell}, {Fazio},
  {Willner}, {Swinbank}, {Arumugam}, {Summers}, {Cohen}, {Jansen}, {Windhorst},
  {Meena}, {Zitrin}, {Keel}, {Coe}, {Conselice}, {D'Silva}, {Driver}, {Frye},
  {Grogin}, {Koekemoer}, {Marshall}, {Nonino}, {Pirzkal}, {Robotham},
  {Rutkowski}, {Ryan}, {Tompkins}, {Willmer}, {Yan}, {Broadhurst}, {Cheng},
  {Diego}, {Kamieneski}, \& {Yun}}]{2023arXiv230616039S}
{Smail}, I., {Dudzeviciute}, U., {Gurwell}, M., {et~al.} 2023, arXiv e-prints,
  arXiv:2306.16039, \dodoi{10.48550/arXiv.2306.16039}

\bibitem[{{Speagle}(2020)}]{2020MNRAS.493.3132S}
{Speagle}, J.~S. 2020, \mnras, 493, 3132, \dodoi{10.1093/mnras/staa278}

\bibitem[{{Speagle} {et~al.}(2014){Speagle}, {Steinhardt}, {Capak}, \&
  {Silverman}}]{2014ApJS..214...15S}
{Speagle}, J.~S., {Steinhardt}, C.~L., {Capak}, P.~L., \& {Silverman}, J.~D.
  2014, \apjs, 214, 15, \dodoi{10.1088/0067-0049/214/2/15}

\bibitem[{{Stach} {et~al.}(2021){Stach}, {Smail}, {Amvrosiadis}, {Swinbank},
  {Dudzevi{\v{c}}i{\={u}}t{\.{e}}}, {Geach}, {Almaini}, {Birkin}, {Chen},
  {Conselice}, {Cooke}, {Coppin}, {Dunlop}, {Farrah}, {Ikarashi}, {Ivison}, \&
  {Wardlow}}]{2021MNRAS.504..172S}
{Stach}, S.~M., {Smail}, I., {Amvrosiadis}, A., {et~al.} 2021, \mnras, 504,
  172, \dodoi{10.1093/mnras/stab714}

\bibitem[{{Stalevski} {et~al.}(2016){Stalevski}, {Ricci}, {Ueda}, {Lira},
  {Fritz}, \& {Baes}}]{2016MNRAS.458.2288S}
{Stalevski}, M., {Ricci}, C., {Ueda}, Y., {et~al.} 2016, \mnras, 458, 2288,
  \dodoi{10.1093/mnras/stw444}

\bibitem[{{Tacconi} {et~al.}(2008){Tacconi}, {Genzel}, {Smail}, {Neri},
  {Chapman}, {Ivison}, {Blain}, {Cox}, {Omont}, {Bertoldi}, {Greve},
  {F{\"o}rster Schreiber}, {Genel}, {Lutz}, {Swinbank}, {Shapley}, {Erb},
  {Cimatti}, {Daddi}, \& {Baker}}]{2008ApJ...680..246T}
{Tacconi}, L.~J., {Genzel}, R., {Smail}, I., {et~al.} 2008, \apj, 680, 246,
  \dodoi{10.1086/587168}

\bibitem[{{Tamura} {et~al.}(2014){Tamura}, {Saito}, {Tsuru}, {Uchida}, {Iono},
  {Yun}, {Espada}, \& {Kawabe}}]{2014ApJ...781L..39T}
{Tamura}, Y., {Saito}, T., {Tsuru}, T.~G., {et~al.} 2014, \apjl, 781, L39,
  \dodoi{10.1088/2041-8205/781/2/L39}

\bibitem[{{Tsukui} {et~al.}(2023){Tsukui}, {Wisnioski}, {Bland-Hawthorn},
  {Mai}, {Iguchi}, {Baba}, \& {Freeman}}]{2023arXiv230814798T}
{Tsukui}, T., {Wisnioski}, E., {Bland-Hawthorn}, J., {et~al.} 2023, arXiv
  e-prints, arXiv:2308.14798, \dodoi{10.48550/arXiv.2308.14798}

\bibitem[{Virtanen {et~al.}(2020)Virtanen, Gommers, Oliphant, Haberland, Reddy,
  Cournapeau, Burovski, Peterson, Weckesser, Bright, {van der Walt}, Brett,
  Wilson, Millman, Mayorov, Nelson, Jones, Kern, Larson, Carey, Polat, Feng,
  Moore, {VanderPlas}, Laxalde, Perktold, Cimrman, Henriksen, Quintero, Harris,
  Archibald, Ribeiro, Pedregosa, {van Mulbregt}, \& {SciPy 1.0
  Contributors}}]{2020SciPy-NMeth}
Virtanen, P., Gommers, R., Oliphant, T.~E., {et~al.} 2020, Nature Methods, 17,
  261, \dodoi{10.1038/s41592-019-0686-2}

\bibitem[{{Wang} {et~al.}(2020){Wang}, {Athanassoula}, {Yu}, {Wolf}, {Shao},
  {Gao}, \& {Randriamampandry}}]{2020ApJ...893...19W}
{Wang}, J., {Athanassoula}, E., {Yu}, S.-Y., {et~al.} 2020, \apj, 893, 19,
  \dodoi{10.3847/1538-4357/ab7fad}

\bibitem[{{Yu} \& {Ho}(2020)}]{2020ApJ...900..150Y}
{Yu}, S.-Y., \& {Ho}, L.~C. 2020, \apj, 900, 150,
  \dodoi{10.3847/1538-4357/abac5b}

\bibitem[{{Yu} {et~al.}(2022{\natexlab{a}}){Yu}, {Xu}, {Ho}, {Wang}, \&
  {Kao}}]{2022A&A...661A..98Y}
{Yu}, S.-Y., {Xu}, D., {Ho}, L.~C., {Wang}, J., \& {Kao}, W.-B.
  2022{\natexlab{a}}, \aap, 661, A98, \dodoi{10.1051/0004-6361/202142533}

\bibitem[{{Yu} {et~al.}(2022{\natexlab{b}}){Yu}, {Kalinova}, {Colombo},
  {Bolatto}, {Wong}, {Levy}, {Villanueva}, {S{\'a}nchez}, {Ho}, {Vogel},
  {Teuben}, \& {Rubio}}]{2022A&A...666A.175Y}
{Yu}, S.-Y., {Kalinova}, V., {Colombo}, D., {et~al.} 2022{\natexlab{b}}, \aap,
  666, A175, \dodoi{10.1051/0004-6361/202244306}

\end{thebibliography}

\
\end{document}